\PassOptionsToPackage{showwarnings}{silence}
\documentclass[9pt,twocolumn,twoside]{opticajnl}
\journal{opticajournal}
\setboolean{shortarticle}{false}
\setboolean{displaycopyright}{false}
\dates{}
\doi{}
\usepackage{amsthm}
\usepackage{braket}
\usepackage{bm}
\usepackage{multirow}
\usepackage{array}
\usepackage{balance}
\usepackage{soul}
\soulregister{\cite}{7}
\soulregister{\ref}{7}
\theoremstyle{plain}

\newtheorem{lemma}{Lemma}

\theoremstyle{definition}

\providecommand{\bibcommenthead}{}

\renewcommand{\bibliographystyle}[1]{}
\setcitestyle{super,open={},close={}}
\hypersetup{hypertexnames=false,colorlinks=true,citecolor=blue,urlcolor=blue,linkcolor=blue}
\makeatletter
\xpatchcmd{\@maketitle}{\vskip50pt}{\vskip20pt}{}{}
\let\arxivAppendixAbscontent\abscontent
\xpatchcmd{\abscontent}{\absfont \theabstract}{\absfont \theabstract\par}{}{}
\makeatother

\AtBeginEnvironment{thebibliography}{\interlinepenalty=10000}
\title{Chip-Based Discrete-Modulated Continuous-Variable Quantum Key Distribution with Uncertainty-Principle-Based Composable Security}

\author[1,$\dagger$]{\mbox{Jiale Mi}}
\author[1,$\dagger$]{\mbox{Yiming Bian}}
\author[2,*]{\mbox{Zhengyu Li}}
\author[1]{\mbox{Xuesong Xu}}
\author[1]{\mbox{Song Yu}}
\author[1,4,*]{\mbox{Yichen Zhang}}
\author[3,*]{\mbox{Hong Guo}}
\affil[1]{State Key Laboratory of Information Photonics and Optical Communications, School of Electronic Engineering, Beijing University of Posts and Telecommunications, Beijing 100876, China}
\affil[2]{Huawei Technologies Co., Ltd., Shenzhen 518129, China}
\affil[3]{Beijing Key Laboratory of Quantum Sensing and Precision Measurement, Center for Quantum Information Technology, and Institute of Quantum Electronics, Peking University, Beijing 100871, China}
\affil[4]{Beijing Key Laboratory of Optical Quantum Information Network, Beijing University of Posts and Telecommunications, Beijing 100876, China}
\affil[*]{Corresponding authors: lizhengyu2@huawei.com; zhangyc@bupt.edu.cn; hongguo@pku.edu.cn}
\affil[$\dagger$]{These authors contributed equally to this work.}

\begin{abstract}
    Continuous-variable quantum key distribution offers a promising platform for scalable quantum networks through its compatibility with standard telecommunications technology and photonic integration. High-order discrete modulation facilitates practical implementation while approaching the key-rate performance of ideal Gaussian modulation, 
    but its finite-size composable security analysis entails substantial computational overhead.
    Here, we develop an uncertainty-principle-based security framework that enables finite-size composable key-rate evaluation against collective attacks for constellations up to 256-QAM.
    By introducing a heterodyne-measured ancillary mode into the entanglement-based source description, we incorporate additional source correlations into the security analysis that allow the Heisenberg uncertainty principle alone to provide a tight bound for the unknown correlations in a covariance matrix characterizing the whole system.
    Matrix symmetrization further reduces the key-rate optimization to a one-dimensional search, while subsequent mode reduction yields an analytical lower bound requiring no numerical search.
    Experimental demonstrations using 256-QAM on discrete-component and silicon-photonic-receiver platforms achieve finite-size composable secret key rates of 17.436 and 0.519 Mbps, respectively, over a 50 km fiber link.
    This approach facilitates the practical deployment of high-speed, real-time and compact quantum key distribution systems.
    Beyond quantum key distribution, the principle of tightening uncertainty-constrained bounds via source-mode expansion offers a paradigm for exploring the information-theoretic properties of complex non-Gaussian systems.

\end{abstract}

\modulolinenumbers[1]

\usepackage{dcolumn,bbding,mathrsfs,cuted,tikz,enumitem,lipsum,mwe}
\usetikzlibrary{matrix,fit}

\makeatletter
\let\arxivSavedMakeTitle\@maketitle
\let\arxivSavedAuthor\@author
\newcommand{\arxivAppendixTitle}{%
  \gdef\@title{Appendix for ``Chip-Based Discrete-Modulated Continuous-Variable Quantum Key Distribution with Uncertainty-Principle-Based Composable Security''}%
  \let\@author\arxivSavedAuthor
  \patchcmd{\AB@affillist}{Center for Quantum Information Technology, and Institute}{and Center for Quantum Information Technology, and Institute}{}{}%
  \def\theabstract{}%
  \let\abscontentformatted\@empty
  \patchcmd{\arxivSavedMakeTitle}{\vskip25pt}{\vskip0pt}{}{}%
  \twocolumn[\arxivSavedMakeTitle]%
  \thispagestyle{fancy}%
}
\makeatother
\begin{document}
\let\arxivSavedAddcontentsline\addcontentsline
\renewcommand{\addcontentsline}[3]{\ifstrequal{#1}{toc}{}{\arxivSavedAddcontentsline{#1}{#2}{#3}}}
\nolinenumbers
\maketitle

\section*{Introduction}

Quantum key distribution (QKD) uses quantum mechanical principles to establish information-theoretically secure keys between remote parties \cite{bennet1984quantum,AdvInQC,PTPQKDRMV2020}. Continuous-variable (CV) QKD protocols with coherent measurement exhibit high compatibility with standard telecommunications \cite{weedbrook2012gaussian,Zhang2024Continuous,usenko2026continuous}.
There are two typical modulation schemes for CV protocols: Gaussian modulation \cite{ralph1999continuous,grosshans2002continuous,weedbrook2004quantum} and discrete modulation \cite{leverrier2009unconditional}.
Gaussian modulation provides a benchmark for the secret key rate (SKR) and has demonstrated strong performance in metropolitan-area links \cite{zhang2020long,hajomer2024long,wang2025high}, free-space links \cite{FreeSpace2026long}, and quantum access networks \cite{hajomer2024passive,pan2025high,bian2026approaching}.
Alongside these advances, composable security has become central to practical QKD, ensuring that the generated keys remain secure when used in subsequent cryptographic applications \cite{pirandola2024improved}. 
Meanwhile, integrated photonic receivers offer a route towards compact and scalable implementations \cite{Bian2024Continuous}.
Despite these advances, Gaussian-modulated systems typically operate at GHz-scale repetition frequencies, with further increases constrained by finite hardware resolution, which makes ideal continuous Gaussian modulation impossible.

\begin{table*}[t]
    \centering
    \captionsetup{skip=4pt}
    \caption{\textbf{Comparison of DM-CV-QKD security-analysis methods.}
    Analytical bound denotes a key-rate lower bound requiring no numerical security optimization.
    Gaussian-like performance requires simulated secret-key-rate curves to nearly overlap the Gaussian-modulation benchmark throughout the practical transmission-distance and excess-noise ranges. A check mark denotes a reported demonstration, with the constellation size in parentheses; a cross denotes no such demonstration.
    Experimental constellation identifies the largest constellation experimentally validated in the corresponding security regime.}
    \label{tab:security_frameworks}
    \small
    \setlength{\tabcolsep}{1.2pt}
    \renewcommand{\arraystretch}{1.12}
    \begin{tabular*}{\textwidth}{@{\extracolsep{\fill}}lcccccc@{}}
        \toprule
        \multirow{2}{*}{\shortstack[l]{Security-analysis\\method}}
        & \multicolumn{3}{c}{Asymptotic security}
        & \multicolumn{3}{c}{Composable security} \\
        \cmidrule(lr){2-4}\cmidrule(lr){5-7}
        & \shortstack{\strut Analytical\\\strut bound}
        & \shortstack{\strut Gaussian-like\\\strut performance}
        & \shortstack{\strut Experimental\\\strut constellation}
        & \shortstack{\strut Analytical\\\strut bound}
        & \shortstack{\strut Gaussian-like\\\strut performance}
        & \shortstack{\strut Experimental\\\strut constellation} \\
        \midrule
        \begin{tabular}[c]{@{}l@{}}SDP with a\\linear objective \cite{DMCVLeverrier}\end{tabular}
        & \(\checkmark\) \cite{denys2021explicit}
        & \(\checkmark\) (256-QAM) \cite{denys2021explicit}
        & 256-QAM \cite{pan2022experimental}
        & \(\times\)
        & \(\times\)
        & Not demonstrated \\
        \addlinespace[3pt]
        \begin{tabular}[c]{@{}l@{}}SDP with a\\nonlinear objective \cite{DMCVLinjie}\end{tabular}
        & \(\times\)
        & \(\times\)
        & 16-QAM \cite{pan2025long}
        & \(\times\)
        & \(\times\)
        & 16-QAM \cite{wu2026high} \\
        \addlinespace[3pt]
        \begin{tabular}[c]{@{}l@{}}Uncertainty principle\\(This work)\end{tabular}
        & \(\checkmark\)
        & \(\checkmark\) (256-QAM)
        & 256-QAM
        & \(\checkmark\)
        & \(\checkmark\) (256-QAM)
        & 256-QAM \\
        \bottomrule
    \end{tabular*}
\end{table*}

Discrete modulation provides a practical alternative in high-speed systems under this hardware constraint.
Higher-order constellations can approach the SKR of Gaussian modulation, but their non-Gaussian structure complicates security analysis \cite{li2018user}.
There are two major frameworks for establishing asymptotic SKR bounds for discrete-modulated (DM) CV-QKD: the linear semidefinite-programming (SDP) approach introduced by Ghorai et al.\ \cite{DMCVLeverrier} and the nonlinear SDP approach developed by Lin et al.\ \cite{DMCVLinjie}. Denys et al.\ subsequently derived explicit analytical asymptotic bounds for arbitrary modulation, substantially reducing the computational burden for large constellations \cite{denys2021explicit}.
Practical security, however, requires finite-size composable bounds that account for statistical fluctuations, error-correction leakage and privacy amplification. Progress in this direction includes the treatment of finite-range heterodyne measurements \cite{lupo2022nonideal}, numerical security proofs incorporating energy and acceptance tests \cite{kanitschar2023finite}, and experimental demonstrations with QPSK and probability-shaped 16-QAM \cite{hajomer2025composable,wu2026high}.
More recently, a QPSK experiment demonstrated finite-size composable security against general attacks with real-time post-processing \cite{bodenstedt2026realtime}.
Nevertheless, efficiently evaluating composable key rates for higher-order constellations remains challenging.
As summarized in Table~\ref{tab:security_frameworks}, existing composable frameworks have been demonstrated for QPSK and 16-QAM, while their computational cost hinders extension to larger constellations, limiting the practical exploitation of higher-order modulation to approach the Gaussian-modulation benchmark. This motivates an efficient analytical framework for finite-size composable SKR evaluation.

\begin{figure*}[t]
    \centering
    \includegraphics[width=0.9\linewidth]{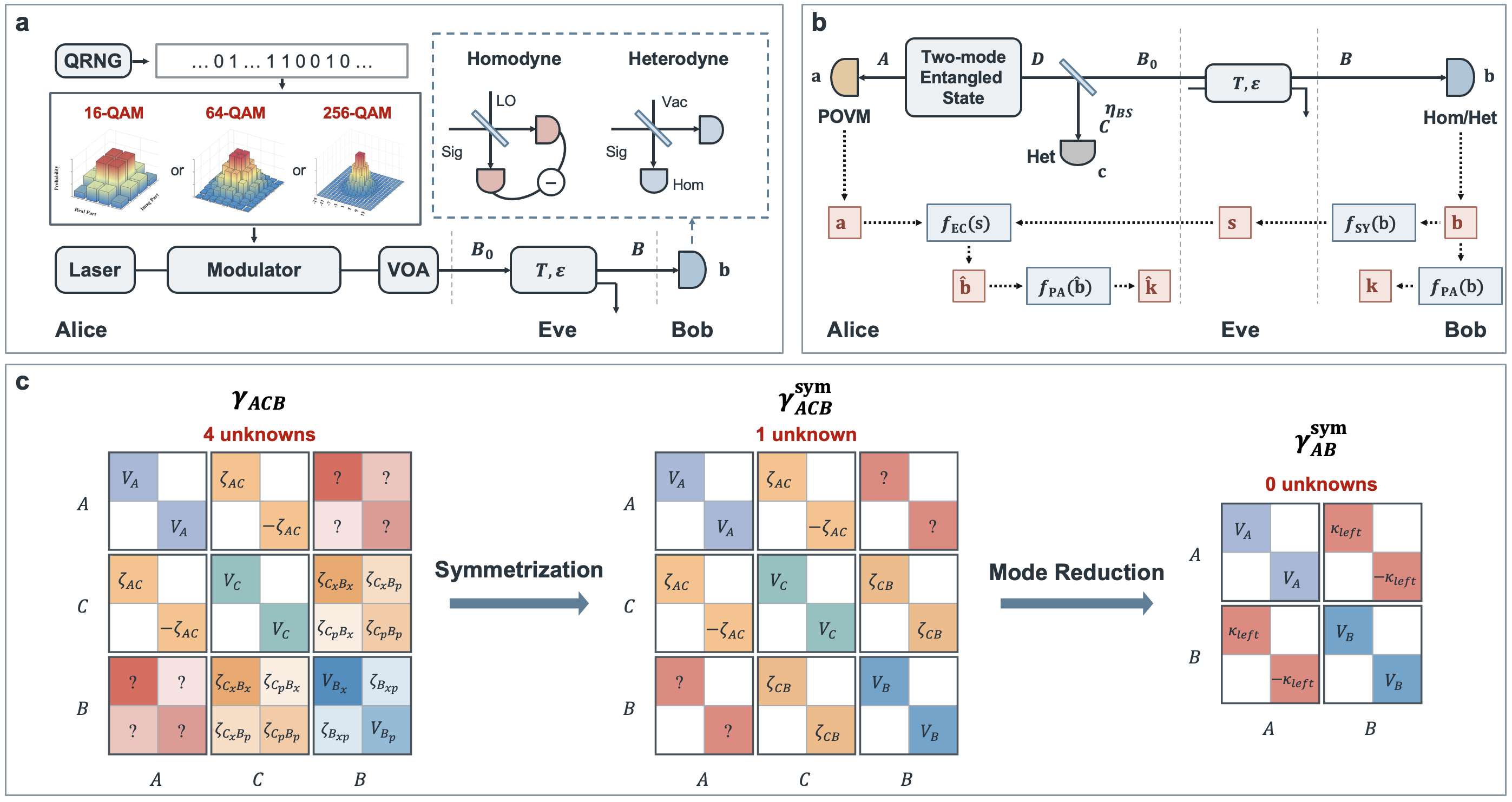}
    \caption{\textbf{Protocol and security framework.}
    \textbf{a} The prepare-and-measure scheme of the proposed DM CV-QKD protocol. A quantum random number generator (QRNG) produces \(q\)-bit symbols, each of which is mapped to one of $2^q$ coherent states in the discrete modulation constellation. Inset plots show the probability distributions of three representative modulation formats: 16-QAM, 64-QAM, and 256-QAM. VOA, variable optical attenuator.
    \textbf{b} The entanglement-based scheme and classical post-processing of the proposed protocol. Bob calculates the syndrome $\mathbf{s}=f_{\mathrm{SY}}(\mathbf{b})$ and sends it to Alice. Using her data $\mathbf{a}$, Alice applies the decoder $f_{\mathrm{EC}}(\mathbf{s})$ to reconstruct $\hat{\mathbf{b}}$. After correctness verification, the same privacy-amplification function produces $\mathbf{k}=f_{\mathrm{PA}}(\mathbf{b})$ and $\hat{\mathbf{k}}=f_{\mathrm{PA}}(\hat{\mathbf{b}})$.
    POVM, positive-operator valued measure. Hom/Het, homodyne/heterodyne measurement.
    \textbf{c} Covariance matrix symmetrization and mode reduction. The white boxes represent 0.
    }\label{ProtocolEB}
\end{figure*}


Alongside advances in security analysis, photonic integration has brought DM CV-QKD closer to compact, scalable implementations. Integrated silicon photonic transmitters and receivers have been demonstrated using 8-PSK modulation over 50.4 km of optical fiber \cite{liu2025integrated}. High-speed chip-based systems have also achieved 16 GBaud over 20 km \cite{hajomer2026chip} and 40 GBaud over 10 km, the latter with finite-size composable security \cite{ng2026gigabit}.
However, extending chip-based DM CV-QKD to longer fiber links with finite-size composable security requires further improvements in key-rate performance. Higher-order constellations offer a promising route, but realizing their potential requires an efficient composable security analysis framework.

In this work, we develop an uncertainty-principle-based framework providing analytical lower bounds on finite-size composable key rates for high-order DM CV-QKD.
By introducing a heterodyne-measured ancillary mode into the entanglement-based source description, we make additional source correlations available for security analysis. The Heisenberg uncertainty principle alone bounds the unknown correlations in the system's covariance matrix, defining a feasible set for worst-case key-rate evaluation. Exploiting the convexity of the SKR, we symmetrize this matrix and reduce the optimization to a one-dimensional search over an analytically determined interval. Further mode reduction yields an analytical lower bound requiring no numerical search.
Incorporating these bounds into a finite-size composable analysis enables high-order key-rate evaluation and supports online estimation.
Simulation results show that the performance of the proposed protocol approaches that of ideal Gaussian modulation with no more than 256 constellation points. 
Compared with the SDP-based methods in Table~\ref{tab:security_frameworks}, our framework provides analytical finite-size composable SKR bounds and supports real-time SKR evaluation for arbitrary constellations, enabling the performance advantages of higher-order modulation to be assessed efficiently.
We validate the framework through 256-QAM experiments on both discrete-component and chip-based CV-QKD platforms, achieving finite-size composable secret key rates of 17.436 and 0.519 Mbps, respectively, over 50 km of optical fiber.
The framework provides a basis for real-time composable key-rate estimation for high-order modulation and supports compact, scalable CV-QKD implementations.
Moreover, the proposed framework provides an explicit physical picture of how source correlations and measured statistics constrain estimates of an eavesdropper's information under non-Gaussian modulation.

\section*{Results}
\subsection*{Protocol and security framework}
The prepare-and-measure (PM) scheme of the proposed protocol is illustrated in Fig. \ref{ProtocolEB}a.
Alice prepares a coherent state $\rho_{B_0}$, randomly chosen from an $M$-coherent-state set \(\{\alpha_{1},\alpha_{2},\dots,\alpha_{M}\}\) with discrete probability distribution \(\{p_1,p_2,\dots,p_M\}\). Alice records the preparation label $\mathrm{a}$, sends the state to Bob, and records the mean of quadratures for \(\rho_{B_0}\), as $(\hat{x}_{B_0}, \hat{p}_{B_0})$. Bob receives the state $\rho_{B}$ and performs homodyne or heterodyne measurement, obtaining the results $\mathrm{b}$, as $(\hat{x}_{B}, \hat{p}_{B})$. 

Building equivalent entanglement-based (EB) scheme is necessary for security analysis. While if we follow the traditional two-mode entangled source $(\rho_{AB_0})$ modeling method, preparing mode $B_0$ in one of a finite constellation of coherent states requires Alice to perform a positive-operator valued measure (POVM) with finite elements on mode $A$, rather than a heterodyne measurement, leaving the quadrature correlations between Alice and Bob inaccessible to direct estimation. The uncertainty principle alone only admits a broad range of unknown correlations, potentially yielding no positive key rate under worst-case evaluation. SDP-based methods obtain tighter bounds by optimizing over quantum states consistent with the known preparation ensemble and observed statistics, but become computationally demanding for large constellations.
To avoid these complex optimizations, we extend the EB source description to a multimode source on Alice’s side, whose additional correlations tighten the bound derived from the uncertainty principle.

We follow the intuition of QKD security originating from the monogamy of entanglement, namely that the stronger the quantifiable correlation between Alice and Bob, the less information is leaked to Eve. 
Specifically, we extend the conventional two-mode EB source $\rho_{AB_0}$ to a three-mode source $\rho_{ACB_0}$ by introducing an ancillary mode $C$. Mode $C$ is heterodyne measured, yielding the outcome $c$ and enabling estimation of the correlations between modes $C$ and $B$. The correlations between modes $A$ and $C$ are determined by the source model. Together with Bob's measured statistics, these quantities determine all blocks of the covariance matrix $\gamma_{ACB}$ except the $A$--$B$ correlation block. The additional source correlations tighten the uncertainty-principle constraints on this unknown block, allowing an upper bound $\chi_{BE}^{G}(\gamma_{ACB})$ on Eve's Holevo information to be evaluated using Gaussian optimality \cite{wolf2006extremality}. $\gamma_{ACB}$ takes the form
\begin{equation} 
    \gamma_{ACB}=
    \begin{pmatrix} 
            \gamma_A & \psi _{AC} & \kappa_{AB} \\ 
    \psi _{AC}^T & \gamma_C & \psi _{CB}\\
    \kappa_{AB}^T & \psi _{CB}^T &\gamma_B
    \end{pmatrix},
\end{equation} 
where each term $\gamma_{ij}:=\frac{1}{2}\left\langle\{\Delta\hat{r}_i,\Delta\hat{r}_j\}\right\rangle$, and $\Delta\hat{r}_i=\hat{r}_i-\langle\hat{r}_i\rangle$, with $\hat{\mathbf{r}}=(\hat{x}_A,\hat{p}_A,\hat{x}_C,\hat{p}_C,\hat{x}_B,\hat{p}_B)^T$, $\langle\hat{r}_i\rangle=\operatorname{Tr}(\hat{r}_i \rho_{ACB})$.
Here, $\gamma_A$, $\gamma_C$, and $\psi _{AC}$ can be theoretically calculated based on the entangled source. $\psi _{CB}$ and $\gamma_B$ can be estimated through the measured data. Only $\kappa_{AB} =(\kappa_{11}\ \kappa_{12}\ ;\ \kappa_{21}\ \kappa_{22})$ is unknown due to the POVM measurement, rendering $\chi_{BE}$ a function of $\kappa_{AB}$. Nevertheless, the uncertainty principle \cite{weedbrook2012gaussian} imposes a constraint on $\gamma_{ACB}$, namely $\gamma_{ACB}+i\Omega \ge 0$,
which restricts the value of $\kappa_{AB}$ within a set $\mathcal{F}_{\kappa_{AB}}$. Therefore, we can find the maximum information leakage $\chi_{BE}(\kappa_{AB}^{\mathbf{worst}})$ by traversing $\mathcal{F}_{\kappa_{AB}}$, and the SKR is defined as
\begin{equation} \label{Eq_SKR}
    K=\beta I(\mathrm{a}:\mathrm{b})-\sup_{\kappa_{AB} \in \mathcal{F}_{\kappa_{AB}}}\chi_{BE}^G(\kappa_{AB}).
\end{equation} 

We develop a simple and effective EB scheme, in which the state sent to Bob is only decided by Alice's POVM result $\mathrm{a}$. This means the sub-state $\rho_{CB_0}^{\mathrm{a}}$ conditioned on $\mathrm{a}$ is a product state, $\rho_{CB_0}^{\mathrm{a}}=\rho_{C}^{\mathrm{a}}\otimes\rho_{B_0}^{\mathrm{a}}$ \cite{Supplemental}. So the mean values of $\rho_C^{\mathrm{a}}$ can be designed to linearly depend on $\rho_{B_0}^{\mathrm{a}}$. Following this way, one only needs to find a two-mode entangled state $|\Psi\rangle_{AD}$ and then let mode $D$ pass through a beam-splitter with transmittance $\eta_{BS}$, yielding modes $C$ and $B_0$, as shown in Fig. \ref{ProtocolEB}b. 
Here, the state of mode $B_0$ is exactly the same state $\rho_{B_0}=\sum_{i=1}^{M}p_i|\alpha_i\rangle\langle\alpha_i|$ prepared in the PM scheme. Considering the beam-splitter transformation from mode $D$ to modes $C$ and $B_0$, the corresponding state of mode $D$ is $\rho_D=\sum_{i=1}^{M}p_i|\beta_i\rangle\langle\beta_i|$, where $|\beta_i\rangle=|\alpha_i/\sqrt{\eta_{BS}}\rangle$. A Schmidt purification of $\rho_D$ then gives
$|\Psi\rangle_{AD}=\sum_{i=1}^{M}\sqrt{\nu_i}|\varphi_i\rangle_A|\theta_i\rangle_D$,
where $\nu_i$ is the eigenvalue of $\rho_D$, and $|\varphi_i\rangle_A$ and $|\theta_i\rangle_D$ are orthogonal states that diagonalize $\rho_A$ and $\rho_D$, respectively. 
Therefore, $\rho_{B_0}$ is specified by the source $|\Psi\rangle_{AD}$ and $\eta_{BS}$. 
Then $\rho_{B_0}$ is transmitted through the quantum channel to Bob, who performs heterodyne detection on the received state to obtain the outcome $\mathrm{b}$. Subsequently, based on the syndrome $\mathbf{s}=f_{\mathrm{SY}}(\mathbf{b})$ sent by Bob, Alice performs reverse reconciliation using her data $\mathbf{a}$ to reconstruct Bob's data, yielding $\hat{\mathbf{b}}$. 
Under successful error correction probability $p_{\mathrm{ec}}$, we have $\hat{\mathbf{b}}=\mathbf{b}$. Finally, Alice and Bob perform privacy amplification to extract secret keys. 

Returning to the PM model, mode \(C\) need not be physically prepared. Since \(\rho_{CB_0}^{\mathrm{a}}=\rho_C^{\mathrm{a}}\otimes\rho_{B_0}^{\mathrm{a}}\), estimating the correlation block \(\psi_{CB}\) requires only the quadrature means of mode \(C\) \cite{Supplemental}, together with Bob’s measured data. These mean values are directly determined by the preparation data of mode \(B_0\) through the beam-splitter relation,
$(\hat{x}_C, \hat{p}_{C})=\sqrt{(1-\eta_{BS})/\eta_{BS}}(\hat{x}_{B_0}, \hat{p}_{B_0})$.
Therefore, the PM scheme has no change in hardware comparing to the existing one-way CV-QKD system. 

We next introduce the numerical and analytical methods for SKR calculation. Since the direct evaluation of SKR with Eq. \ref{Eq_SKR} requires an exhaustive search over four unknowns of $\kappa_{AB}$, as shown in the first matrix in Fig. \ref{ProtocolEB}c, it suffers from a significant computational burden. To simplify the searching process, we symmetrize $\gamma_{ACB}$ into the symmetric form \(\gamma_{ACB}^{\mathrm{sym}}\), which includes only one unknown parameter $\kappa$, as the second matrix in Fig. \ref{ProtocolEB}c.
Suppose the initial three-mode quantum state is $\rho_{ACB}^{(0)}$ with covariance matrix 
    \begin{equation} 
    \gamma_{ACB}^{(0)}=\gamma_{ACB}=
    \begin{pmatrix} 
    V_{A}I_{2} & \phi_{AC}\,\sigma_{Z} & \kappa_{AB}^{(0)} \\ 
    \phi_{AC}\,\sigma_{Z} & V_CI_{2} & \psi _{CB}^{(0)}\\
    (\kappa_{AB}^{(0)})^T & (\psi _{CB}^{(0)})^T &\gamma_B^{(0)}
    \end{pmatrix},
\end{equation} 
where the upper-left $4\times 4$ sub-matrix $\gamma_{AC}$ is already symmetric.
The remaining sub-matrices are $\kappa_{AB}^{(0)}=(\kappa_{11}\ \kappa_{12}\ ;\ \kappa_{21}\ \kappa_{22})$, $\psi _{CB}^{(0)}=(\phi_{11}\ \phi_{12}\ ;\ \phi_{21}\ \phi_{22})$, $\gamma_B^{(0)}=(b_{11}\ b_{12}\ ;\ b_{12}\ b_{22})$. Then there exists a state $\rho_{ACB}^{\mathrm{sym}}$, whose covariance matrix takes the symmetric form
\begin{equation}
\gamma_{ACB}^{\mathrm{sym}}=
\begin{pmatrix}
V_A I_2 & \phi_{AC}\sigma_Z & \kappa \sigma_Z \\
\phi_{AC}\sigma_Z & V_C I_2 & \phi I_2 \\
\kappa \sigma_Z & \phi I_2 & V_B I_2
\end{pmatrix},
\end{equation}
where $\phi=\sqrt{(\frac{\phi_{11}+\phi_{22}}{2})^2+(\frac{\phi_{12}-\phi_{21}}{2})^2}$, $\kappa=\frac{(\kappa_{11}-\kappa_{22})(\phi_{11}+\phi_{22})}{4\phi}
+\frac{(\kappa_{12}+\kappa_{21})(\phi_{12}-\phi_{21})}{4\phi}$, $V_B=(b_{11}+b_{22})/2$. 
The corresponding SKRs satisfy $K\big(\rho_{ACB}^{\mathrm{sym}}\big) \le K\big(\rho^{(0)}_{ACB}\big)$. 
The symmetrization consists of three steps detailed in Methods, with the proof of the key-rate inequality based on SKR convexity provided in the Appendix \cite{Supplemental}.

Regarding the analytical boundary, solving the uncertainty-principle constraint, $\gamma_{ACB}^{\mathrm{sym}}+i\Omega \ge 0$, requires all principal minors to be non-negative. The sixth-order principal minor gives the tightest constraint on $\kappa$ as $\kappa \in \mathcal{F}_{\kappa}=[\bar{\kappa}-R, \bar{\kappa}+R]$, where 
\begin{subequations}
    \begin{align}
        \bar{\kappa} &= \sqrt{T\eta_{A}\,\eta_{BS}\big(V_{A}^{2}-1\big)},\\
R &= \sqrt{T\varepsilon(1-\eta_{A})(V_{A}+1)}.
    \end{align}
\end{subequations}
Here, $\eta_A=\phi_{AD}^2/(V_A^2-1)$ is a dimensionless parameter to quantify the deviation between $|\Psi\rangle_{AD}$ and an ideal two-mode squeezed-vacuum state $|\Psi\rangle_{AA_0}$ with the same $V_A$ \cite{Supplemental}, $\phi_{AD}$ represents the covariance of mode $A$ and $D$.
$T$ and $\varepsilon$ are the equivalent channel transmittance and input-referred excess noise.
Finally, Eq. \ref{Eq_SKR} can be updated as
\begin{equation}
\label{eq:key_rate_3mode}
    K_{\mathrm{3-mode}}=\beta I(\mathrm{a}:\mathrm{b})-\sup_{\kappa \in [\bar{\kappa}-R, \bar{\kappa}+R]}\chi_{BE}^G(\kappa),
\end{equation}
yielding an SKR lower bound calculated numerically from the three-mode covariance matrix.


\begin{figure}[t]
    \centering
    \includegraphics[width=\linewidth]{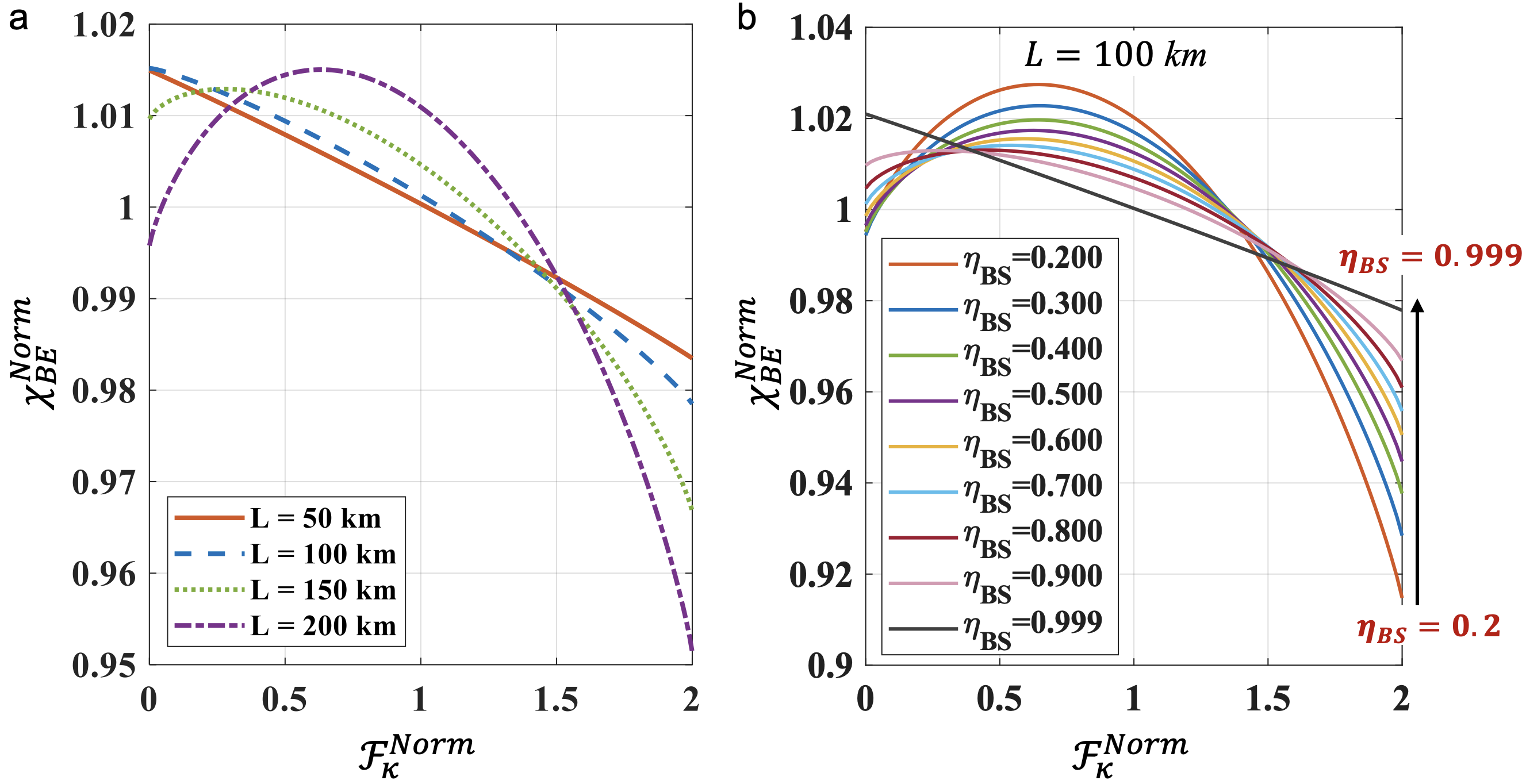}
    \caption{\textbf{Holevo information over the feasible correlation interval.}
    \textbf{a} For different transmission distances $L$. \textbf{b} For different $\eta_{BS}$ at 100 km.
    The variation of normalized $\chi_{BE}^{\mathrm{Norm}}$ within the normalized interval $\mathcal{F}_{\kappa}^{\mathrm{Norm}}$, where $\tilde{\kappa}=1+\frac{\kappa-\bar{\kappa}}{R}\in\mathcal{F}_{\kappa}^{\mathrm{Norm}}$ denotes the normalized parameter within $\mathcal{F}_{\kappa}^{\mathrm{Norm}}$.
    Simulation parameters: excess noise $\varepsilon=0.04$ SNU, detection efficiency $\eta=0.6$, electronic noise $\nu_{ele}=0.1$ SNU. 
    The modulation variance $V_M=4$ SNU, and $\eta_A=1-10^{-4}$.
    }\label{SBE}
\end{figure}

Numerically, a bisection search over the interval $\mathcal{F}_{\kappa}$ typically requires about six iterations to locate the true $\kappa_t$ and the corresponding maximum $\chi_{BE}^{G}(\kappa_t)$, with a relative difference of less than $0.1\%$ between the SKRs obtained in two consecutive iterations.

For the analytical lower bound of SKR, one may intuitively expect it to occur at the left boundary $\kappa_l=\bar{\kappa}-R$ of \(\mathcal{F}_{\kappa}\), since weaker correlations between Alice and Bob generally increase Eve's accessible information and thus reduce the SKR. However, this intuition does not always hold for the three-mode covariance matrix. 
As shown in Fig. \ref{SBE} (a), \(\chi_{BE}^{\mathrm{Norm}}\), which is obtained by normalizing \(\chi_{BE}\) with its mean value, decreases monotonically over the normalized interval $\mathcal{F}_{\kappa}^{\mathrm{Norm}}$ at short distances ($L=50$ and 100 km). At long distances ($L=150$ and 200 km), it exhibits a nonmonotonic trend, first increasing and then decreasing.
Fig. \ref{SBE} (b) shows the influence of $\eta_{BS}$ on $\chi_{BE}$ over $\mathcal{F}_{\kappa}^{\mathrm{Norm}}$ at a fixed transmission distance of 100 km, where the modulation variance is set as $V_M=4$ SNU, and $V_A=V_M/\eta_{BS}+1$. 
As $\eta_{BS}$ increases, the variation of $\chi_{BE}^{\mathrm{Norm}}$ becomes smoother and more monotonic, and the relative deviation between the true worst-case point and the left boundary \(\kappa_{\mathrm{left}}\) roughly decreases. In the limit $\eta_{BS} \to 1$, the worst-case point approaches the left boundary, indicating that the characteristics of the three-mode matrix gradually converge to those of a two-mode matrix.
To gain insight into this phenomenon, we apply an entanglement attack model \cite{Supplemental}, from which we infer that Eve's state becomes nontrivial and entangled near the left physical boundary, potentially weakening Eve's ability and resulting in the nonlinear behavior.

To obtain a rigorous analytical lower bound on the SKR, we first determine the admissible interval of $\kappa$ from $\gamma_{ACB}^{\mathrm{sym}}$, then discard mode $C$ to obtain $\gamma_{AB}^{\mathrm{sym}}$, as shown in the third matrix in Fig. \ref{ProtocolEB}c. This is conservative because assigning the discarded mode to Eve cannot decrease her Holevo information. Specifically, strong subadditivity implies $S(C|E)\geq S(C|E\mathrm{b})$, where $\mathrm{b}$ denotes Bob's continuous classical measurement outcome. Using the definition of Holevo information, we obtain $\chi(\mathrm{b}:CE)-\chi(\mathrm{b}:E)=S(C|E)-S(C|E\mathrm{b})=I(\mathrm{b}:C|E)\geq 0$,
With the two-mode matrix, we can thus evaluate the analytical key-rate lower bound directly at the left boundary $\kappa_l=\bar{\kappa}-R$:
\begin{equation}
\label{eq:key_rate_2mode}
    K_{\mathrm{2-mode}}=\beta I(\mathrm{a}:\mathrm{b})-\chi_{BE}^{G}(\bar{\kappa}-R).
\end{equation}
Here, $\chi_{BE}^{G}$ is the Gaussian Holevo bound evaluated from $\gamma_{AB}^{\mathrm{sym}}$, with the discarded mode $C$ assigned to Eve. This expression requires no numerical search over $\kappa$.
This may come at the cost of a reduced key rate, with less than $2\%$ \cite{Supplemental} degradation for distances below 150 km when $\eta_{BS}$ approaches 1, whereas a more noticeable performance loss is observed beyond 150 km.

\subsection*{Composable key rate}

Composable security requires Alice and Bob to obtain identical keys, and the key should be uniformly random and independent of Eve. 
For key-rate evaluation, we set $m=n=N$, allowing all retained samples to be used for both parameter estimation (PE) and key extraction by performing information reconciliation before PE \cite{wang2019allrawkeys}.
For these signals, $\mathbf{a}=(\mathrm{a}_1,\ldots,\mathrm{a}_n)$ contains Alice's labels for the transmitted coherent states, which are her POVM outcomes in the EB scheme. Bob discretizes the corresponding measurement outcomes $\mathrm{b}_i$ into symbols $l_i\in\mathcal{L}$, forming his raw-key sequence $\mathbf{b}=(l_1,\ldots,l_n)$. He then calculates the syndrome $\mathbf{s}=f_{\mathrm{SY}}(\mathbf{b})$ and sends it to Alice. Using her data $\mathbf{a}$, Alice applies $f_{\mathrm{EC}}(\mathbf{s})$ to recover $\hat{\mathbf{b}}$. Alice and Bob compare verification hashes and discard the block if they disagree. The correctness parameter $\epsilon_{\mathrm{cor}}$ bounds the probability that different sequences pass this test. Let $p_{\mathrm{ec}}$ denote the probability that reconciliation and verification succeed.

For each accepted block, Alice and Bob apply the same randomly selected two-universal hash function to obtain $\mathbf{k}=f_{\mathrm{PA}}(\mathbf{b})$ and $\hat{\mathbf{k}}=f_{\mathrm{PA}}(\hat{\mathbf{b}})$. Secrecy requires $\frac{p_{\mathrm{ec}}}{2}\left\|\rho_{KEP\mid\mathrm{pass}}-\tau_K\otimes\rho_{EP\mid\mathrm{pass}}\right\|_1\leq\epsilon_{\mathrm{sec}}$, where $K$ is Bob's final key register, $E$ is Eve's system, $P$ is the register containing all public information, and $\tau_K$ is the uniform key state. 
Excluding parameter-estimation failures, we set $\epsilon_{\mathrm{sec}}=\epsilon_s+\epsilon_h$, where $\epsilon_s$ is the smoothing parameter and $\epsilon_h$ is the privacy-amplification error parameter.

After successful reconciliation and verification, the accepted data are described by a subnormalized state $\sigma^n$, with $\operatorname{Tr}\sigma^n=p_{\mathrm{ec}}$, which need not have a product form. Tensor-product reduction allows us to bound its $\epsilon_s$-smooth min-entropy from below using the original product state $\rho_{lE}^{\otimes n}$ \cite{pirandola2024improved},
\begin{equation}
\label{eq:composable_min_entropy}
H_{\min}^{\epsilon_s}(l^n|E^n)_{\sigma^n}
\geq H_{\min}^{\epsilon_s}(l^n|E^n)_{\rho_{lE}^{\otimes n}}.
\end{equation}
Using the asymptotic equipartition property (AEP) for the product state in \eqref{eq:composable_min_entropy}, and including reconciliation leakage, hash verification, privacy amplification, and the statistical fluctuations in PE, we obtain the composable key rate per transmitted signal
\begin{equation}
\label{eq:composable_key_rate}
K_{\mathrm{com}}^{\mathrm{3-mode}}
=\frac{p_{\mathrm{ec}}}{N}
\left[nR_{\mathrm{wc}}^{\mathrm{3-mode}}-\sqrt{n}\,\Delta_{\mathrm{AEP}}
+\log_2(\epsilon_h^2\epsilon_{\mathrm{cor}})\right],
\end{equation}
Here $\Delta_{\mathrm{AEP}}=4\log_2(\sqrt{|\mathcal{L}|}+2)\sqrt{\log_2(2/\epsilon_s^2)}$ quantifies the finite-size effect associated with the AEP, $R_{\mathrm{wc}}^{\mathrm{3-mode}}$ is the worst-case SKR calculated from the three-mode covariance matrix with PE uncertainty, as calculated below. 

To calculate $R_{\mathrm{wc}}^{\mathrm{3-mode}}$, we use $m$ samples to obtain the original 3-mode covariance matrix, and then apply the symmetrization procedure to determine the corresponding values of the equivalent channel parameters $T$ and $\varepsilon$. For a phase-insensitive linear channel with Gaussian noise, the corresponding confidence intervals under the large-sample Gaussian approximation are
\begin{equation}
\label{eq:pe_T_interval}
T\in\mathcal{I}_T
=\left[\hat{T}-w\sigma_T,\,\hat{T}+w\sigma_T\right],
\end{equation}
\begin{equation}
\label{eq:pe_noise_interval}
\varepsilon\in\mathcal{I}_{\varepsilon}
=\Bigg[
\frac{\hat{T}\hat{\varepsilon}}{\hat{T}+w\sigma_T}-w\sigma_{\varepsilon},\frac{\hat{T}\hat{\varepsilon}}{\hat{T}-w\sigma_T}+w\sigma_{\varepsilon}
\Bigg],
\end{equation}
where
\begin{align}
\sigma_T&=\frac{2\hat{T}}{\sqrt{V_0m}}
\sqrt{c_{\mathrm{pe}}+
\frac{\hat{\varepsilon}+(V_0+v_{\mathrm{el}})/(\eta\hat{T})}{V_M}},\\
\sigma_{\varepsilon}&=\sqrt{\frac{2}{V_0m}}\,
\frac{\eta\hat{T}\hat{\varepsilon}+V_0+v_{\mathrm{el}}}
{\eta(\hat{T}-w\sigma_T)}.
\end{align}
Here $V_M$ is Alice's modulation variance, $\eta$ and $v_{\mathrm{el}}$ are the calibrated detector efficiency and electronic-noise variance, and $V_0=1$ or $2$ for homodyne or heterodyne detection. The coefficient $c_{\mathrm{pe}}$ is set to zero. The noise range includes the uncertainty in $T$, with $\sigma_{\varepsilon}$ evaluated using its lower confidence limit to give a conservative range. The parameter $w$ is given by $w=\sqrt{2}\,\operatorname{erf}^{-1}(1-\epsilon_{\mathrm{pe}})$. Here, $\epsilon_{\mathrm{pe}}$ is the failure probability assigned to each parameter estimate, and the total PE failure probability is at most $n_{\mathrm{pm}}\epsilon_{\mathrm{pe}}$. 
Then the unknown correlations in the matrix $\gamma_{ACB}^{\mathrm{sym}}(T,\varepsilon,\kappa)$ are specified by the single parameter $\kappa\in\mathcal{F}_{\kappa}(T,\varepsilon)=[\bar{\kappa}(T,\varepsilon)-R(T,\varepsilon),\bar{\kappa}(T,\varepsilon)+R(T,\varepsilon)]$. 
Therefore, the resulting Holevo information bound is calculated as
\begin{equation}
\label{eq:composable_holevo_wc}
\chi_{\mathrm{wc}}^{\mathrm{3-mode}}
=\sup_{\substack{T\in\mathcal{I}_T,\;
\varepsilon\in\mathcal{I}_{\varepsilon},\;\kappa\in\mathcal{F}_{\kappa}(T,\varepsilon)}}
\chi_{BE}^{G}\!\left(\gamma_{ACB}^{\mathrm{sym}}(T,\varepsilon,\kappa)\right).
\end{equation}
Combining this Holevo bound with the estimated Alice--Bob correlation gives $R_{\mathrm{wc}}^{\mathrm{3-mode}}=\beta I(\mathrm{a}:l)-\chi_{\mathrm{wc}}^{\mathrm{3-mode}}$, which is used directly in \eqref{eq:composable_key_rate}. Here, $\beta I(\mathrm{a}:l)\leq H(l)-\mathrm{leak}_{\mathrm{ec}}/n$ accounting for the actual syndrome leakage. 

\begin{figure}[t]
    \centering
    \includegraphics[width=\linewidth]{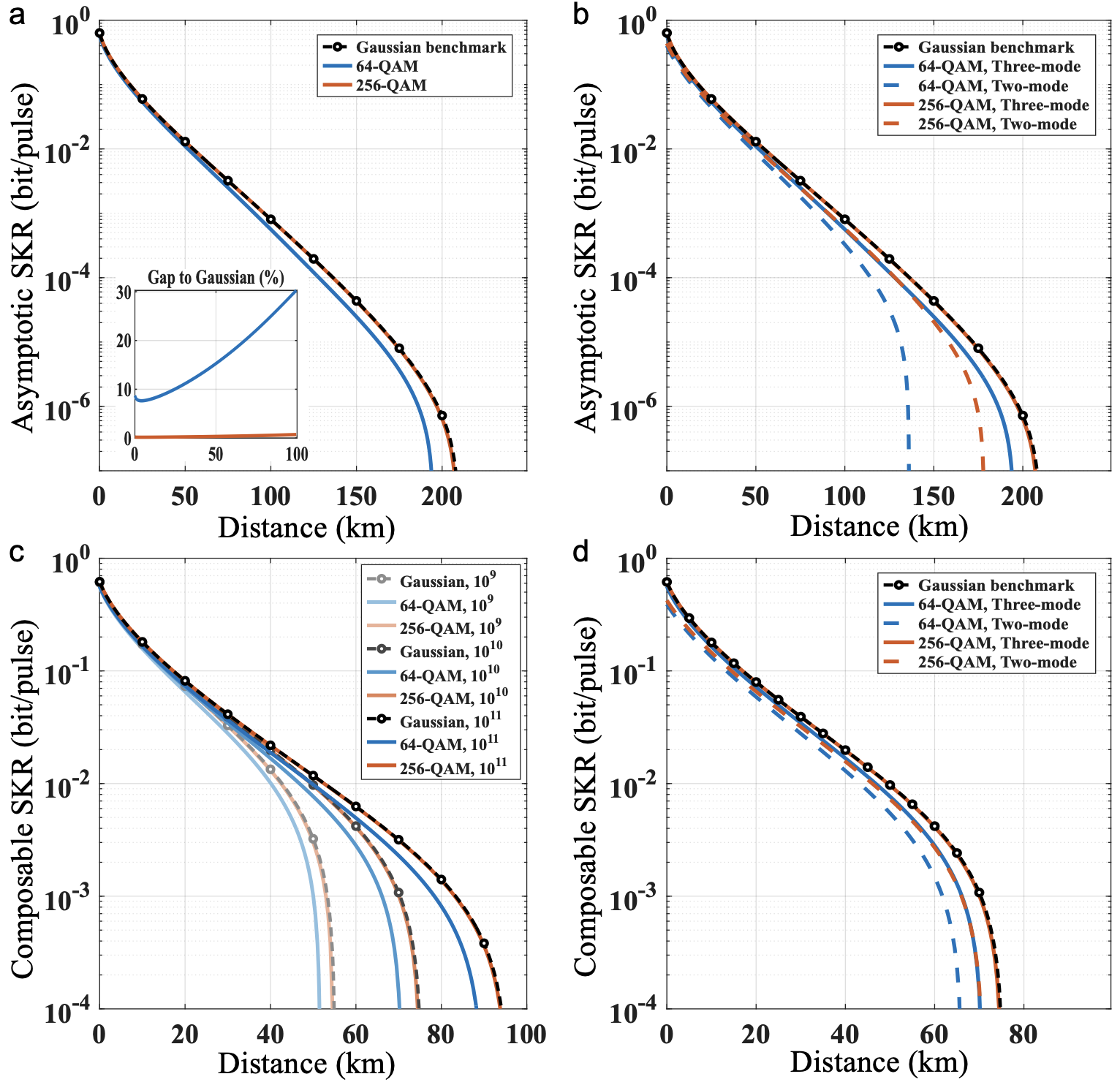}

    \caption{\textbf{Asymptotic and composable key-rate performance.}
    \textbf{a} Asymptotic SKRs from three-mode $\gamma_{ACB}^{\mathrm{sym}}$ compared with Gaussian modulation. The inset shows the relative gap to the Gaussian benchmark.
    \textbf{b} Comparison between asymptotic SKRs from three-mode $\gamma_{ACB}^{\mathrm{sym}}$ and two-mode $\gamma_{AB}^{\mathrm{sym}}$.
    \textbf{c} Composable SKRs from three-mode $\gamma_{ACB}^{\mathrm{sym}}$ for $N=10^9$, $10^{10}$, and $10^{11}$.
    \textbf{d} Comparison between composable SKRs from three-mode $\gamma_{ACB}^{\mathrm{sym}}$ and two-mode $\gamma_{AB}^{\mathrm{sym}}$ at $N=10^{10}$.
    Common parameters: source-mode variance $V_A=5$ SNU, modulation variance $V_M=3.6$ SNU, $\eta_{BS}=0.9$, $\eta_A=1-1.35\times10^{-4}$ for 64-QAM and $1-6.5\times10^{-8}$ for 256-QAM, reconciliation efficiency $\beta=0.95$, excess noise $\varepsilon=0.04$ SNU, detection efficiency $\eta=0.6$, and electronic noise $\nu_{ele}=0.1$ SNU. For \textbf{c} and \textbf{d}, $p_{\mathrm{ec}}=0.98$, $\epsilon_s=\epsilon_h=\epsilon_{\mathrm{cor}}=\epsilon_{\mathrm{pe}}=2^{-32}$, $|\mathcal{L}|=2^{16}$.
    }\label{SimulationPerformance}
\end{figure}
\begin{figure*}[t]
    \includegraphics[width=\linewidth]{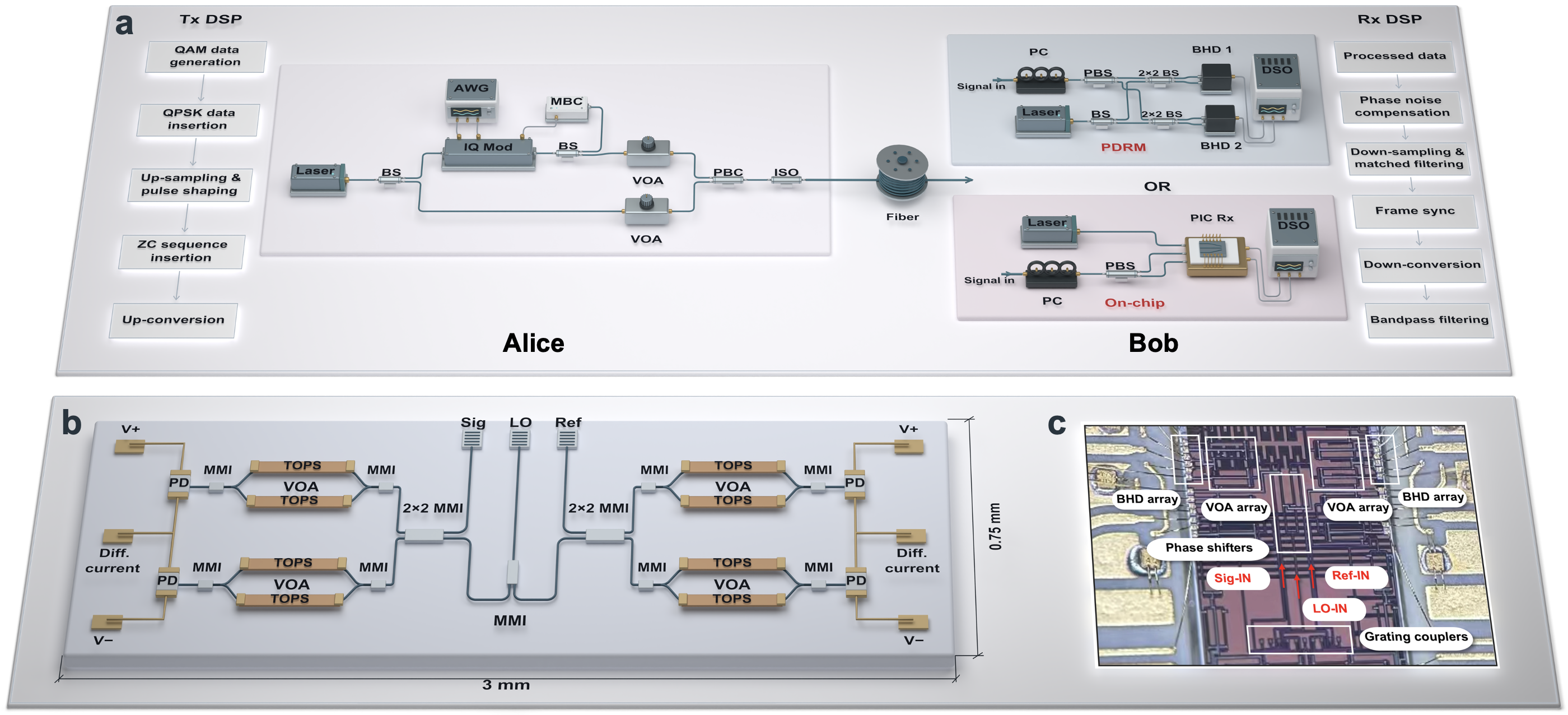}
    \caption{\textbf{Experimental setup and integrated receiver.}\label{Fig: Exp_Scheme}
    \textbf{a} Experimental setup of the DM CV-QKD system and DSP routine. AWG, arbitrary waveform generator; IQ Mod, in-phase/quadrature modulator; VOA, variable optical attenuator; PC, polarization controller; PBC, polarization beam combiner; PBS, polarization beam splitter; BS, beam splitter; PMOC, polarization-maintaining optical couplers; BHD, balanced homodyne detector; PDRM, polarization diversity receiver module; DSO, digital storage oscilloscope.
    \textbf{b} The integrated silicon photonic receiver chip, which replaces the PDRM. PD, photodiode; MMI, multi-mode interferometer. 
    \textbf{c} Microscope photo of the chip.
    }
\end{figure*}

Furthermore, if we give mode $C$ to Eve and use the reduced two-mode matrix $\gamma_{AB}^{\mathrm{sym}}$, the worst-case channel parameters are the smallest transmittance and largest excess noise, namely $T_{\mathrm{wc}}=\min\mathcal{I}_T=\hat{T}-w\sigma_T$ and $\varepsilon_{\mathrm{wc}}=\max\mathcal{I}_{\varepsilon}=\hat{T}\hat{\varepsilon}/(\hat{T}-w\sigma_T)+w\sigma_{\varepsilon}$. The correlation is evaluated directly at $\kappa_l^{\mathrm{wc}}=\bar{\kappa}(T_{\mathrm{wc}},\varepsilon_{\mathrm{wc}})-R(T_{\mathrm{wc}},\varepsilon_{\mathrm{wc}})$, giving
\begin{equation}
\label{eq:composable_holevo_2mode}
\chi_{\mathrm{wc}}^{\mathrm{2-mode}}
=\chi_{BE}^{G}\!\left(
\gamma_{AB}^{\mathrm{sym}}(T_{\mathrm{wc}},\varepsilon_{\mathrm{wc}},\kappa_l^{\mathrm{wc}})
\right).
\end{equation}
The corresponding composable key rate is
\begin{equation}
\label{eq:composable_key_rate_2mode}
K_{\mathrm{com}}^{\mathrm{2-mode}}
=\frac{p_{\mathrm{ec}}}{N}\Big[
nR_{\mathrm{wc}}^{\mathrm{2-mode}}-\sqrt{n}\,\Delta_{\mathrm{AEP}}+\log_2(\epsilon_h^2\epsilon_{\mathrm{cor}})\Big],
\end{equation}
where $R_{\mathrm{wc}}^{\mathrm{2-mode}}=\beta I(\mathrm{a}:l)-\chi_{\mathrm{wc}}^{\mathrm{2-mode}}$. The SKR calculated from the two-mode covariance matrix is therefore obtained directly, without searching over $\kappa$, $T$, or $\varepsilon$. The PE contribution is weighted by the probability $p_{\mathrm{ec}}$ that a block passes reconciliation. Both rates therefore have the total security parameter
\begin{equation}
\label{eq:composable_security_budget}
\epsilon=\epsilon_{\mathrm{cor}}+\epsilon_s+\epsilon_h
+p_{\mathrm{ec}}n_{\mathrm{pm}}\epsilon_{\mathrm{pe}},
\end{equation}
with $n_{\mathrm{pm}}=2$ for the transmittance and excess-noise estimates.

\subsection*{Simulations}
Numerical simulations assess how closely high-order discrete modulation approaches the Gaussian key-rate benchmark and quantify the performance cost of the analytical bound calculated from the two-mode covariance matrix. We use a phase-insensitive linear channel with Gaussian noise and fiber attenuation of $0.2$ dB/km. Both QAM constellations and the Gaussian reference have the same transmitted modulation variance, $V_M=3.6$ SNU. Moreover, $\phi_{CB}$ and $V_B$ can be calculated as $\phi_{CB}=\sqrt{T\eta_{BS}(1-\eta_{BS})}(V_{A}-1)$, $V_B=T(\eta_{BS}(V_A-1)+\varepsilon)+1$. 
Figure~\ref{SimulationPerformance}a shows that the asymptotic SKR of 256-QAM calculated from the three-mode covariance matrix nearly approaches the Gaussian benchmark, while 64-QAM gives a lower SKR and shorter transmission range. The inset resolves their relative gaps, reflecting the source parameter $\eta_A$ being closer to unity for 256-QAM. Fig. \ref{SimulationPerformance}b compares the bounds calculated from the two-mode and three-mode covariance matrices at $\eta_{BS}=0.9$. Discarding mode $C$ gives an analytical SKR bound without numerical optimization, but reduces the SKR and transmission range for both constellations.
Fig.~\ref{SimulationPerformance}c shows that increasing $N$ from $10^9$ to $10^{11}$ reduces finite-size effects and improves both the composable SKR and transmission range. For every block size, 256-QAM remains closer to the Gaussian benchmark than 64-QAM. Fig.~\ref{SimulationPerformance}d compares the composable bounds calculated from the two-mode and three-mode covariance matrices at $N=10^{10}$ under identical security settings. The bound calculated from the two-mode covariance matrix offers a simpler calculation with a lower SKR, and its relative performance loss becomes more pronounced near the transmission limit.

\begin{figure*}[t]
    \centering
    \includegraphics[width=0.98\linewidth]{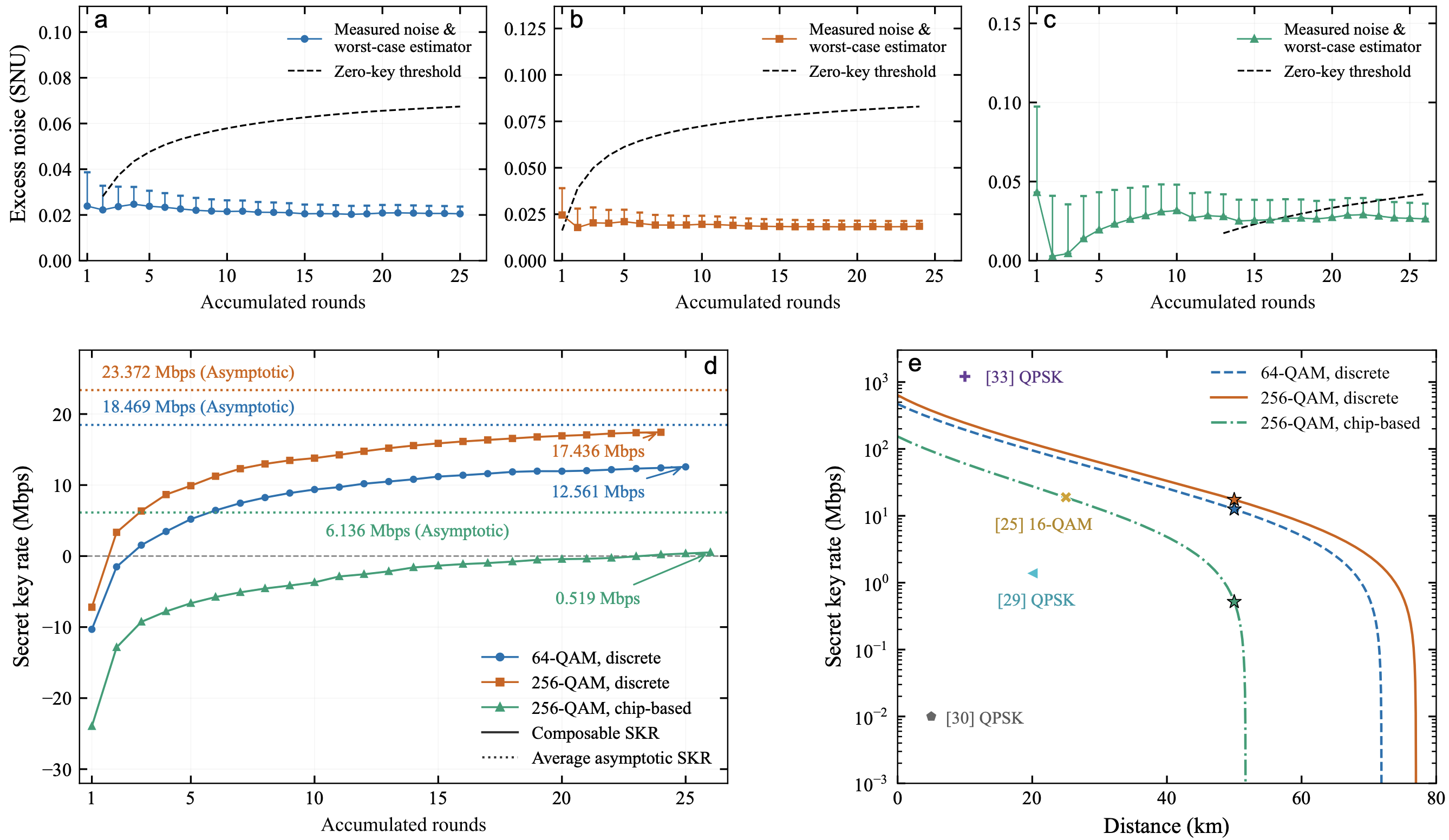}
    \caption{\textbf{Experimental results.}
    \textbf{a--c} Excess-noise estimates as the number of accumulated rounds increases for discrete-component 64-QAM, discrete-component 256-QAM, and chip-based 256-QAM at 50 km, respectively. Each round contains $2.56\times10^8$ samples. Points show the excess noise estimated from the accumulated data, and one-sided error bars extend to its worst-case estimator. Black dashed lines indicate the corresponding zero-key thresholds.
    \textbf{d} Composable SKRs calculated from the cumulative experimental parameters. Blue, orange, and green correspond to the configurations in \textbf{a--c}. Dotted lines show asymptotic SKRs at the final mean parameters.
    \textbf{e} The simulated composable SKRs for the discrete-component and chip-based system, respectively. Stars mark the experimental results at 50 km, and other markers show reported QPSK and 16-QAM results \cite{ng2026gigabit,wu2026high,hajomer2025composable,bodenstedt2026realtime} for comparison.
    }\label{Fig_ExpRes}
\end{figure*}

\subsection*{Experimental demonstration}
The proposed protocol is experimentally demonstrated on both the discrete-module-based and chip-based platforms. The experimental setup of the system and the DSP routine is illustrated in Fig. \ref{Fig: Exp_Scheme}a.

At the transmitter, Alice employs a continuous-wave (CW) laser (NKT Photonics Basik X15) operating at 1550.12 nm with a 0.1 kHz linewidth as the optical carrier. The laser output is split by a beam splitter into two branches: one undergoes QAM modulation, while the other serves as a pilot tone to provide a frequency and phase reference.
The QAM modulation is implemented using an in-phase/quadrature (IQ) modulator (Fujitsu FTM7962EP) driven by an arbitrary waveform generator (AWG) at a sampling rate of 30 GSa/s. With a bandwidth of 23 GHz, the IQ modulator supports a system baud rate of 1.5 GBaud. 
To ensure stability, a portion of the modulated signal is directed to a bias controller that maintains the modulator in carrier-suppression mode. 
Subsequently, a variable optical attenuator (VOA, EXFO FTBx-3500-BI) reduces the signal to the quantum level, achieving an average photon number consistent with the target modulation variance. 
This VOA integrates a real-time power monitor for precise calibration of the output power.
Finally, the modulated quantum signals and pilot tones are merged by a polarization beam combiner (PBC). They are aligned in orthogonal polarization directions and co-propagated through the untrusted quantum channel.

The receiver, Bob, is designed to support two distinct configurations for shot-noise-limited coherent detection: a discrete Polarization-Diversity Receiver Module (PDRM) and an integrated on-chip solution.
In the PDRM configuration, the incoming signal from the untrusted quantum channel passes through a polarization controller (PC) followed by a polarization beam splitter (PBS) to demultiplex the orthogonal polarization components (the quantum signal and the pilot tone). A local oscillator (LO), provided by a dedicated laser (NKT Photonics Basik X15), is split by a 1:2 beam splitter (BS). The LO center frequency is offset by approximately 1.55 GHz relative to Alice’s transmitter laser to enable intermediate-frequency (IF) coherent detection. The resulting signals are then captured and digitized by a digital storage oscilloscope (DSO) with 8-bit resolution for further analysis.
Alternatively, Bob can be implemented using a silicon photonic integrated chip. In this architecture, the signal (after the PC and PBS) and the LO laser are coupled directly into a packaged silicon-photonic receiver, where two shot-noise-limited homodyne detectors and their corresponding balance-control modules are monolithically integrated on a 220-nm silicon-on-insulator (SOI) platform, as illustrated in Fig. \ref{Fig: Exp_Scheme}b. The chip features a compact footprint of $3\times 0.75\ mm^2$, and the microscope photo of the chip is shown in Fig. \ref{Fig: Exp_Scheme}c. The output electrical signals are similarly processed by the DSO. Compared to discrete fiber-based components, this integrated approach significantly minimizes the system footprint while enhancing scalability and mechanical stability.
For simplicity and accuracy, the one-time shot noise unit calibration method is adopted \cite{zhang2020one}.
The detection efficiencies of the PDRM and the silicon photonic receiver are approximately 0.6366 and 0.2197, respectively.
The system performance is ensured by a comprehensive DSP routine, categorized into transmitter-side (Tx) and receiver-side (Rx) processing.

Figure~\ref{Fig_ExpRes} connects the measured excess noise statistics to finite-size key generation over 50 km. Fig.~\ref{Fig_ExpRes}a--c correspond to discrete-component 64-QAM, discrete-component 256-QAM, and chip-based 256-QAM, comprising 25, 24, and 26 rounds, respectively. Each round contributes $2.56\times10^8$ samples, so the final cumulative sample sizes are $N=6.400\times10^9$, $6.144\times10^9$, and $6.656\times10^9$. 
The channel transmittance and excess noise are then re-estimated from this enlarged data set. The final excess-noise estimates are 0.020503, 0.018520, and 0.026416 SNU after all rounds are accumulated. It can be seen that the statistical correction generally narrows as rounds accumulate. In the chip-based configuration, the zero-key threshold appears only from round 13, at smaller sample sizes, even zero nominal excess noise does not yield a positive rate.

Figure~\ref{Fig_ExpRes}d shows the resulting composable SKRs calculated from the three-mode covariance matrix with trusted detector modelling. For SKR estimation, the reconciliation efficiency is $\beta=0.96$, and the error-correction success probability is $p_{\mathrm{ec}}=0.9$ for all three configurations. The system operates at a 1.5 GBaud rate with a training sequence ratio of 1/8.  At the final accumulated rounds, the discrete-component 64-QAM and 256-QAM configurations yield 12.561 and 17.436 Mbps, respectively, while chip-based 256-QAM yields approximately 0.519 Mbps. The average asymptotic reference rates are 18.469, 23.372, and 6.136 Mbps, respectively. The separation from these references shows the remaining finite-size effect, which is particularly consequential for the integrated receiver.

The experimental composable key rates are shown in Fig.~\ref{Fig_ExpRes}e. The blue dashed, orange solid, and green dash-dotted curves represent the simulated SKRs for discrete-component 64-QAM, discrete-component 256-QAM, and chip-based 256-QAM systems, respectively. The experimental key rates at 50 km are marked with stars and compared with reported composable results using QPSK and 16-QAM \cite{ng2026gigabit,wu2026high,hajomer2025composable,bodenstedt2026realtime}.
Although the integrated photonic receiver exhibits higher loss, it still enables effective key generation with high-order constellations.

\section*{Discussion}

In this work, we have developed a security analysis framework that enables analytical evaluation of finite-size composable key-rate calculation for high-order DM CV-QKD, providing a basis for real-time secret key-rate estimation. The framework uses source correlations and measured covariance-matrix elements to bound Eve's information through the Heisenberg uncertainty principle. Covariance-matrix symmetrization reduces the search over four unknown correlation parameters to a one-dimensional optimization, while further mode reduction gives an analytical bound without numerical search. Incorporating parameter estimation, error correction, and privacy amplification then allows the composable rate to be calculated directly from the reduced matrix. This analytical calculation avoids semidefinite programming and repeated numerical optimization, making the framework suitable for high-order constellations and online operation.

The framework combines efficient key-rate evaluation with high performance. The bound calculated from the three-mode matrix approaches the ideal Gaussian-modulation benchmark in asymptotic simulations, while the analytical bound calculated from the two-mode matrix offers a simpler calculation with a conservative rate. The composable simulations show that 256-QAM remains closer to the Gaussian reference than 64-QAM after finite-size effects are included. We experimentally demonstrate the framework using a telecom-compatible transmitter and both discrete-component and silicon photonic receivers. Using the bound calculated from the three-mode covariance matrix, the discrete-component system achieves composable SKRs of 12.561 and 17.436 Mbps for 64-QAM and 256-QAM, respectively, over 50 km of optical fiber, while the chip-based system achieves 0.519 Mbps with 256-QAM at the same distance. These results demonstrate high-order modulation with composable security and integrated detection within the same security-analysis framework.

For practical operation, source parameters can be precomputed for a fixed configuration and reused as channel conditions change, allowing analytical key-rate updates from the measured statistics. The physical description also relates constellation geometry and preparation probabilities to the achievable rate, providing guidance for source optimization and extension to other modulation formats. Combining analytical rate estimation with real-time error correction and privacy amplification would support continuous key generation with composable security. Improvements in receiver coupling loss and detection efficiency, together with optimized block sizes and parameter estimation, could further increase the key rates and transmission distances. Extending the framework to multi-user quantum access networks offers a direction toward compact, scalable quantum communication systems with efficient security evaluation.

\section*{Methods}
\subsection*{Covariance matrix symmetrization}

We detail the symmetrization steps for covariance matrix as shown in Fig. \ref{Fig_Sym}. Firstly, suppose the initial three-mode quantum state is $\rho_{ACB}^0$, with covariance matrix $\gamma_{ACB}^{(0)}\ (=\gamma_{ACB})$, which can be written as
\begin{equation} 
    \gamma_{ACB}^{(0)}=\gamma_{ACB}=
    \begin{pmatrix} 
    V_{A}I_{2} & \phi_{AC}\,\sigma_{Z} & \kappa_{AB}^{(0)} \\ 
    \phi_{AC}\,\sigma_{Z} & V_CI_{2} & \psi _{CB}^{(0)}\\
    (\kappa_{AB}^{(0)})^T & (\psi _{CB}^{(0)})^T &\gamma_B^{(0)}
    \end{pmatrix},
\end{equation} 
where the upper-left $4\times 4$ sub-matrix $\gamma_{AC}$ is symmetric due to the intrinsic symmetry of the designed source, where 
\begin{equation}
    \begin{aligned}
    V_C=&(1-\eta_{BS})(V_A-1)+1, \\
    \phi _{AC}=&\sqrt{(1-\eta_{BS})\eta_A(V_A^2-1)},
    \end{aligned}
\end{equation}
and $I_2= (1\ 0\ ;\ 0\ 1)$, $\sigma_Z=(1\ 0\ ;\ 0\ -1)$. The remaining sub-matrices are estimated using the measurement data, resulting in an unknown matrix $\kappa_{AB}$, and two non-symmetric matrices $\psi_{CB}$ and $\gamma_{B}$,
\begin{equation}
    \begin{aligned}
    \kappa_{AB}^{(0)}&=(\kappa_{11}\ \kappa_{12}\ ;\ \kappa_{21}\ \kappa_{22}),\\
    \psi _{CB}^{(0)}&=(\phi_{11}\ \phi_{12}\ ;\ \phi_{21}\ \phi_{22}),\\
    \gamma_B^{(0)}&=(b_{11}\ b_{12}\ ;\ b_{12}\ b_{22}).
    \end{aligned}
\end{equation}
Then \(\gamma_{ACB}^{(0)}\) is symmetrized by the three steps as below:

\begin{figure}[t]
    \centering
    \includegraphics[width=\linewidth]{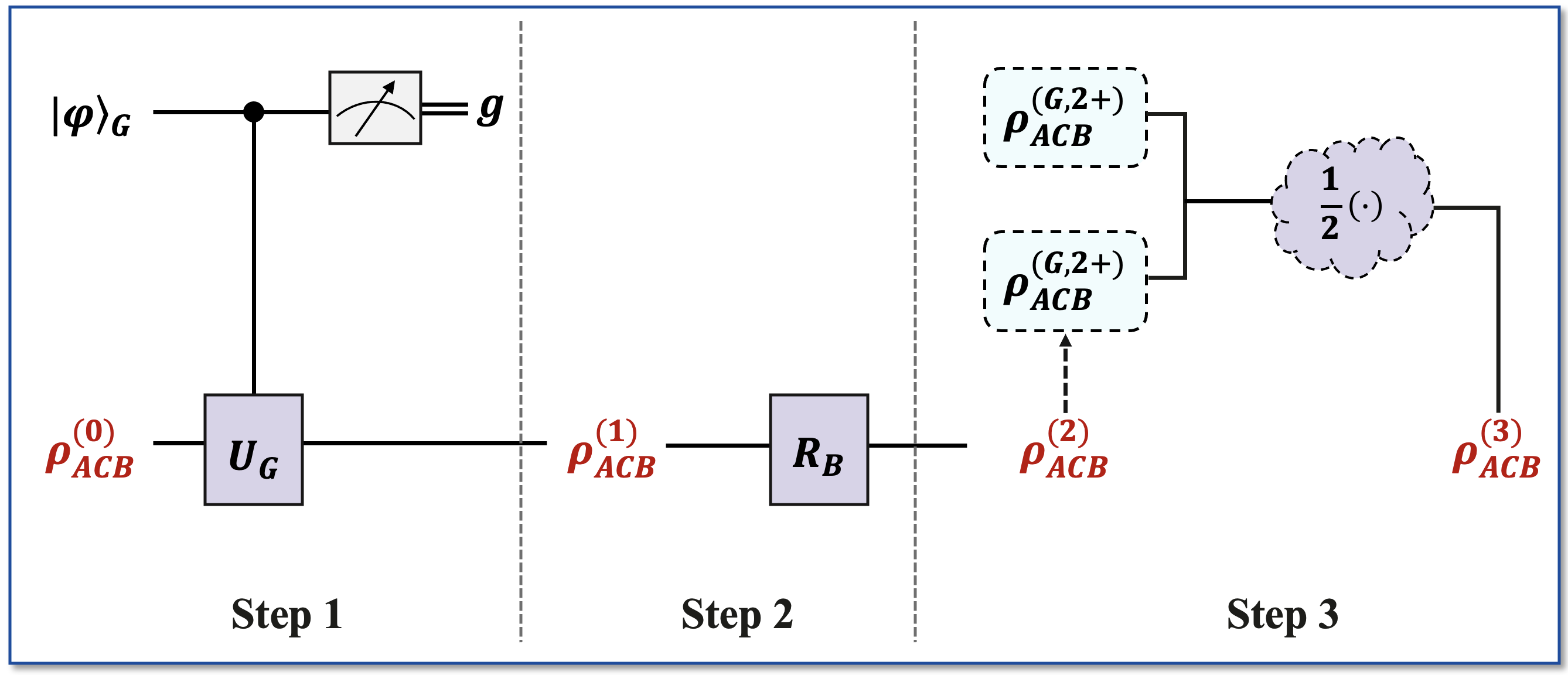}
    \caption{\textbf{Covariance matrix symmetrization.}
    \textbf{Step 1}: Controlled rotation. $U_G$ is a controlled operation applied to $\rho_{ACB}^{(0)}$. \textbf{Step 2}: Local rotation. $R_B$ is a local rotation operation applied on mode $B$. \textbf{Step 3}: State mixture. The two states are mixed according to $\big(\rho_{ACB}^{(G,2+)}+\rho_{ACB}^{(G,2-)}\big)\big/2$.
    }\label{Fig_Sym}
\end{figure}

\textbf{Step 1:}\textit{ Controlled rotation.}
Introduce an auxiliary system $G$ as a control quantum state $|\varphi \rangle_G = \sqrt{1/2}(|0 \rangle_G+|1 \rangle_G)$, whose controlled operation is $U_G=|0\rangle\langle0|_G\otimes I_{ABC} +|1\rangle\langle1|_G\otimes R_A\!\left(\tfrac{\pi}{2}\right)\otimes R_{BC}\!\left(-\tfrac{\pi}{2}\right)$, where $|\varphi \rangle_G=|0\rangle_G$ leaves $\rho_{ACB}^{(0)}$ unchanged, and $|\varphi \rangle_G=|1\rangle_G$ yields the state $\rho_{ACB}^{\mathrm{(0R)}}$ with covariance matrix $\gamma_{ACB}^{\mathrm{(0R)}}$. The resulting equally mixed state and covariance matrix is
\begin{equation}
   \rho_{ACB}^{(1)}=\frac{\rho_{ACB}^{(0)}+\rho_{ACB}^{\mathrm{(0R)}}}{2},\quad \gamma_{ACB}^{(1)}=\frac{\gamma_{ACB}^{(0)}+\gamma_{ACB}^{\mathrm{(0R)}}}{2},
\end{equation} 
where the new sub-matrices are $\gamma_{B}^{(1)}=V_BI_2$ and $V_B=\frac{b_{11}+b_{22}}{2}$, $\kappa_{AB}^{(1)}=(\kappa^{(1)}\ \Delta\kappa^{(1)} \ ;\ \Delta\kappa^{(1)}\ -\kappa^{(1)})$, $\psi _{CB}^{(1)}=(\phi^{(1)}\ \Delta\phi^{(1)} \ ;\ -\Delta\phi^{(1)}\ \phi^{(1)})$, where 
$\kappa^{(1)}=\frac{\kappa_{11}-\kappa_{22}}{2}$, 
$\Delta\kappa^{(1)}=\frac{\kappa_{12}+\kappa_{21}}{2}$, 
$\phi^{(1)}=\frac{\phi_{11}+\phi_{22}}{2}$, 
$\Delta\phi^{(1)}=\frac{\phi_{12}-\phi_{21}}{2}$. 
It's obvious that $K(\rho_{ACB}^{(0)})=K(\rho_{ACB}^{\mathrm{(0R)}})$ because the rotation is a unitary operation. With the convexity of SKR, we have $K\big(\rho_{ACB}^{(1)}\big) \le K\big(\rho^{(0)}_{ACB}\big)$.

\textbf{Step 2:}\textit{ Local rotation.} 
A local rotation $R_B$ with symplectic matrix $X=(\cos \theta\ -\sin \theta\ ;\ \sin \theta\ \cos \theta)$ is applied to mode $B$, yielding $\gamma_{ACB}^{(2)}=(I_{4} \oplus X)\gamma_{ACB}^{(1)}(I_{4} \oplus X)^T$.
Choosing $\tan \theta= -(\Delta\phi^{(1)})/(\phi^{(1)})$, the sub-matrix is symmetrized as $\psi _{CB}^{(2)}=\psi _{CB}^{(1)}X^T=\phi I_2$, and $\phi =\sqrt{(\phi^{(1)})^2 + (\Delta\phi^{(1)})^2}$. 
Similarly, $\kappa _{AB}^{(2)}=\kappa _{AB}^{(1)}X^T=(\kappa^{(2)}\ \Delta\kappa^{(2)}\ ;\ \Delta\kappa^{(2)}\ -\kappa^{(2)})$,
where $\kappa^{(2)}=\kappa^{(1)} \cos \theta - \Delta\kappa^{(1)} \sin \theta$, and $\Delta\kappa^{(2)}=\kappa^{(1)} \sin \theta + \Delta\kappa^{(1)} \cos \theta$.
This unitary operation does not change SKR, therefore we have $K\big(\rho_{ACB}^{(2)}\big) = K\big(\rho_{ACB}^{(1)}\big)$.

\textbf{Step 3:} \textit{State mixture.}
Consider a Gaussian state $\rho_{ACB}^{(G,2+)}$ having covariance matrix $\gamma_{ACB}^{(2+)}$ ($=\gamma_{ACB}^{(2)}$). Let $\rho_{ACB}^{(G,2-)}$ denote a Gaussian state with covariance matrix $\gamma_{ACB}^{(2-)}$, obtained by replacing $\Delta\kappa^{(2)}$ in $\gamma_{ACB}^{(2+)}$ with $-\Delta\kappa^{(2)}$. The equally mixed state and its covariance matrix are
\begin{equation}
    \rho_{ACB}^{\mathrm{sym}}=\frac{\rho_{ACB}^{(G,2+)}+\rho_{ACB}^{(G,2-)}}{2},\quad  \gamma_{ACB}^{\mathrm{sym}}=\frac{\gamma_{ACB}^{(2+)}+\gamma_{ACB}^{\mathrm{(2-)}}}{2},
\end{equation}
where the sub-matrix $\kappa _{AB}^{\mathrm{sym}}=\kappa \sigma_Z$ and $\kappa=\kappa^{(2)}$. Details are provided in Appendix \cite{Supplemental}. Since the symplectic eigenvalues for $\gamma^{(2+)}_{ACB}$ and $\gamma_{ACB}^{(2-)}$ are the same, then $\rho_{ACB}^{(G,2+)}$ and $\rho_{ACB}^{(G,2-)}$ yield the same SKR. Following the convexity of SKR and the Gaussian optimality theorem, $K\big(\rho_{ACB}^{\mathrm{sym}}\big) \le K\big(\rho_{ACB}^{(G,2+)}\big) \le K\big(\rho_{ACB}^{(2)}\big)$.

Ultimately, we obtain a scaling chain of SKR 
\begin{equation}
        K\big(\rho_{ACB}^{\mathrm{sym}}\big) \le K\big(\rho_{ACB}^{(2)}\big)
        =K\big(\rho_{ACB}^{(1)}\big) \le K\big(\rho^{(0)}_{ACB}\big),
\end{equation}
which demonstrates that we can calculate the SKR lower bound with a symmetric matrix, thus reducing the problem to solving for the single parameter $\kappa$.

\subsection{Analytical solution for the range of $\kappa$}\label{kappa}
After we symmetrize the three-mode covariance matrix $\gamma_{ACB}$ and get $\gamma_{ACB}^{\mathrm{sym}}$, we reduce the problem to solving the single parameter $\kappa$. Applying the uncertainty-principle constraint $\gamma_{ACB}^{\mathrm{sym}}+i\Omega \ge 0$, the tightest feasible interval for $\kappa$ is obtained from the sixth-order principal minor, yielding $\kappa \in \mathcal{F}_{\kappa}=[\bar{\kappa}-R, \bar{\kappa}+R]$, where $\bar{\kappa} = \sqrt{T_{cb}\,\eta_{A}\,(1-\eta_{BS})\,\big(V_{A}^{2}-1\big)}$, and $R = \sqrt{T_{cb}\,\varepsilon_{cb}\,(1-\eta_{A})\,(V_{A}+1)}$.
The parameters \(T_{cb}\) and \(\varepsilon_{cb}\) are defined from the symmetrized two-mode covariance matrix \(\gamma_{CB}^{\mathrm{sym}}\) of modes \(B\) and \(C\) to provide an equivalent description of the channel transmittance and excess noise, where
\begin{equation}\label{eq:gamma_CB}
\gamma_{CB}^{\mathrm{sym}}=\begin{pmatrix} V_C I_2 & \phi_{CB} I_2 \\ \phi_{CB} I_2 & V_B I_2 \end{pmatrix}.
\end{equation}

For a matrix having the following form (assuming \(a>b\))
\begin{equation}\label{eq:gamma}
\gamma=\begin{pmatrix} aI_2 & cI_2 \\ cI_2 & bI_2 \end{pmatrix},
\end{equation}
the uncertainty principle requires $\det\bigl(\gamma+i\Omega\bigr)\ge 0$, which can be solved as
\begin{equation}\label{eq:c_upper}
c^2\le (a-1)(b-1)\le (a-1)^2.
\end{equation}
So we can define the channel transmittance and excess noise as
\begin{equation}\label{eq:T_def}
T_{\mathrm{ch}}=\frac{c^2}{(a-1)^2},\quad T_{\mathrm{ch}}\le 1, \quad \varepsilon_{\mathrm{ch}}=\frac{b-1}{T}-(a-1).
\end{equation}

Similarly, for the two-mode covariance matrix \(\gamma_{CB}^{\mathrm{sym}}\), we can define
\begin{equation}\label{eq:Tcb_epscb}
T_{cb}=\frac{\phi_{CB}^2}{(V_C-1)^2},\qquad
\varepsilon_{cb}=\frac{V_B-1}{T_{cb}}-(V_C-1).
\end{equation}
With these two parameters, we can obtain the analytical expression for the range of $\kappa$ as above. 

If we use the channel transmittance $T$ and excess noise $\varepsilon$ defined in Fig. \ref{ProtocolEB}, as a linear channel model for performance simulation, then $\bar{\kappa}$ and $R$ can be re-expressed as $\bar{\kappa} = \sqrt{T\eta_{A}\eta_{BS}\,\big(V_{A}^{2}-1\big)}$, and $R = \sqrt{T\varepsilon(1-\eta_{A})(V_{A}+1)}$. And the correspondence between the parameters under these two definitions is
\begin{equation}
T_{cb}=\frac{\eta_{BS}}{1-\eta_{BS}}T,\qquad
\varepsilon_{cb}=\frac{1-\eta_{BS}}{\eta_{BS}}\varepsilon.
\end{equation}


\subsection*{Compatibility with trusted detector modeling}\label{Compatibility}
To demonstrate the applicability of the proposed symmetrization framework in practical CV-QKD systems, we show that the symmetrization procedure developed for the original three-mode covariance matrix can be naturally extended to the detector-modeling scenario. In particular, the symmetrization of the three-mode state can be consistently generalized to the corresponding five-mode covariance matrix arising from homodyne or heterodyne detection.

For heterodyne detection, we focus on the intermediate-frequency system, in which the \(x\) and \(p\) quadratures can be simultaneously obtained by down-converting the outputs of one balanced homodyne detector. In this case, the heterodyne detector can be considered to consist of two symmetrical homodyne detectors with the same detection efficiency $\eta$ and electronic noise $v_{el}$. Therefore, with the detector modeling, the original three-mode system \(ACB_1\) is extended to the five-mode system \(ACBFG\), and the covariance matrix can be written as 

\begingroup
\setlength{\arraycolsep}{2pt}
\renewcommand{\arraystretch}{0.95}
\begin{equation}
\gamma_{ACBFG}^{(0)}=
\begin{pmatrix}
V_AI_2 & \phi_{AC}\sigma_Z & \kappa_{AB} & \kappa_{AF} & 0\\
\phi_{AC}\sigma_Z & V_CI_2 & \psi_{CB} & \psi_{CF} & 0\\
\kappa_{AB}^T & \psi_{CB}^T & \gamma_B & \psi_{BF} & \phi_{BG}\sigma_Z\\
\kappa_{AF}^T & \psi_{CF}^T & \psi_{BF}^T & \gamma_F & \phi_{FG}\sigma_Z\\
0 & 0 & \phi_{BG}\sigma_Z & \phi_{FG}\sigma_Z & V_GI_2
\end{pmatrix}.
\label{eq:gamma_ACBFG}
\end{equation}
\endgroup
Writing the original sub-matrices as $\kappa_{AB_1}=(\kappa_{ij})$, $\psi_{CB_1}=(\phi_{ij})$, and $\gamma_{B_1}=(b_{ij})$, the detector-model correlation blocks are $\kappa_{AB}=\sqrt{\eta}\,\kappa_{AB_1}$, $\kappa_{AF}=\sqrt{1-\eta}\,\kappa_{AB_1}$, $\psi_{CB}=\sqrt{\eta}\,\psi_{CB_1}$, and $\psi_{CF}=\sqrt{1-\eta}\,\psi_{CB_1}$. The remaining blocks are $\gamma_B=\eta\gamma_{B_1}+(1-\eta)V_GI_2$, $\gamma_F=(1-\eta)\gamma_{B_1}+\eta V_GI_2$, and $\psi_{BF}=\sqrt{\eta(1-\eta)}(\gamma_{B_1}-V_GI_2)$.
Here $V_G=1+v_{el}/(1-\eta)$ for homodyne detection and $V_G=1+2v_{el}/(1-\eta)$ for heterodyne detection.

Now we show that the five-mode covariance matrix can be symmetrized by the modified three-step symmetrization procedure. Since all blocks involving mode \(G\) are already symmetric, they remain unchanged throughout the symmetrization process and will not be discussed further.

In the first step, the controlled \(90^\circ\) phase-space rotation is modified as follows: conditioned on the value of the control state, mode \(A\) is randomly rotated by \(90^\circ\) clockwise, while modes \(C\), \(B\), \(F\), and \(G\) are simultaneously rotated by \(90^\circ\) counterclockwise. After the rotation, the covariance matrices \(\gamma_B\) and \(\gamma_F\) are simultaneously symmetrized. Meanwhile, the correlation blocks \(\phi_{CB}\) and \(\phi_{CF}\) have identical diagonal entries and opposite off-diagonal entries.

In the second step, a local rotation is applied to mode \(B\). Since the covariance matrices \(\phi_{CB}\) and \(\phi_{CF}\) differ only by an overall scaling factor, they share the same rotation angle. Consequently, the same local rotation simultaneously diagonalizes both correlation blocks.
The resulting covariance matrix is $\gamma_{ACBFG}^{(2)}=\mathcal{M}(K)$, where $K=\bigl(\begin{smallmatrix}\kappa&\Delta\kappa\\\Delta\kappa&-\kappa\end{smallmatrix}\bigr)$ and
\begingroup
\setlength{\arraycolsep}{2pt}
\renewcommand{\arraystretch}{0.95}
\begin{equation}
\mathcal{M}(X)=
\begin{pmatrix}
V_AI_2 & \phi_{AC}\sigma_Z & \sqrt{\eta}X & \sqrt{1-\eta}X & 0\\
\phi_{AC}\sigma_Z & V_CI_2 & \phi_{CB}I_2 & \phi_{CF}I_2 & 0\\
\sqrt{\eta}X & \phi_{CB}I_2 & V_BI_2 & \phi_{BF}I_2 & \phi_{BG}\sigma_Z\\
\sqrt{1-\eta}X & \phi_{CF}I_2 & \phi_{BF}I_2 & V_FI_2 & \phi_{FG}\sigma_Z\\
0 & 0 & \phi_{BG}\sigma_Z & \phi_{FG}\sigma_Z & V_GI_2
\end{pmatrix}.
\label{eq:gamma_ACBFG_rot}
\end{equation}
\endgroup

In the third step, the two states with covariance matrices corresponding to opposite values of \(\Delta\kappa\) are mixed with equal probabilities. Direct calculation shows that these two covariance matrices possess identical symplectic eigenvalues and therefore yield the same SKR. Consequently, the mixture satisfies the convexity requirement of the SKR, and the SKR scaling argument established for the three-mode case remains valid.

After these three steps, the symmetric covariance matrix is $\gamma_{ACBFG}^{\mathrm{sym}}=\mathcal{M}(\kappa\sigma_Z).$
Furthermore, the resulting symmetrized covariance matrix is identical to that obtained by first applying the symmetrization procedure to the original three-mode covariance matrix and then extending it to the five-mode matrix with detector modeling. Therefore, the detector model neither alters the symmetrization framework nor affects the SKR scaling relations derived in the previous section. The detector parameters, including the detection efficiency and the electronic noise, remain unchanged throughout the symmetrization procedure.



\balance
\section*{Acknowledgements}
This research was supported by the National Cryptologic Science Fund of China (2025NCSF02050), the National Natural Science Foundation of China (U24B20135), and the Fundamental Research Funds for the Beijing University of Posts and Telecommunications (2026TSQY07).

\section*{Author contributions}
Z.L., Y.Z., and H.G. conceived and supervised the research. J.M. and Y.B. developed the theoretical framework, performed the experiments, and analysed the data. X.X. assisted with the experiments. S.Y. contributed to the research discussions and provided guidance. J.M. and Y.B. wrote the manuscript with input from Z.L., S.Y., Y.Z., and H.G. All authors discussed the results and reviewed the manuscript. J.M. and Y.B. contributed equally to this work.

%
%
%
%
\begingroup
\setlength{\parindent}{0pt}
\setlength{\parskip}{0.65\baselineskip}
\titleformat{name=\section,numberless}{\normalfont\sffamily\bfseries\fontsize{11}{13}\selectfont}{}{0pt}{#1}
\section*{Additional information}
\noindent\textbf{Supplementary information} is provided with this manuscript.

\noindent\textbf{Data availability:} Data underlying the results presented in this paper are not publicly available at this time but may be obtained from the authors upon reasonable request.

\noindent\textbf{Competing interests:} The authors declare no competing interests.
\par
\endgroup

\clearpage
\nobalance
\let\addcontentsline\arxivSavedAddcontentsline
\let\abscontent\arxivAppendixAbscontent
\setcounter{secnumdepth}{2}
\setcounter{section}{0}
\setcounter{subsection}{0}
\setcounter{equation}{0}
\setcounter{figure}{0}
\setcounter{table}{0}
\setcounter{lemma}{0}
\setcounter{theorem}{0}
\setcounter{example}{0}
\renewcommand{\theHsection}{appendix.\arabic{section}}
\renewcommand{\theHequation}{appendix.\arabic{equation}}
\renewcommand{\theHfigure}{appendix.\arabic{figure}}
\renewcommand{\theHtable}{appendix.\arabic{table}}
\resetlinenumber[1]
\renewcommand{\hl}[1]{#1}
\makeatletter\let\sl@StoreMessage\@gobble\makeatother

\nolinenumbers

\arxivAppendixTitle
\tableofcontents

\section{Theoretical methods}
This section provides the theoretical details. Under our framework, it is necessary to introduce ancillary entangled modes into the source and perform heterodyne detection on them to assist the security analysis. We first generalize the source to an $N$-mode entangled source $|\Psi\rangle_{A\mathbf{C}B_0}$ in Sec. \ref{app:Nmode}, where $\mathbf{C}=C_1C_2\dots C_{N-2}\ (N>2)$. 
Then we prove the product feature of the sub-state $\rho_{\mathbf{C}B_0}^a$ conditioned on $a$ in Sec. \ref{app:POVM}. By exploiting this feature, we provide a simple representation of $\rho_{\mathbf{C}}$ in Sec. \ref{app:simply_rhoC}, which removes the need of practical simulation of $c$ in the PM scheme. 
Then in Sec. \ref{app:SourceDesign}, we explain the design for three-mode entangled source and present the optimal results for source parameters under quadrature-amplitude modulation (QAM) cases. 
For the security analysis, we first derive the SKR formula based on the covariance matrix in Sec. \ref{app:SKR}, then we prove the convexity of SKR in Sec. \ref{app:Convexity}. 
The analytical solution for the range of $\kappa$ is solved in Sec. \ref{app:kappa}.
With this property, we accomplish the symmetrization of the 3-mode covariance matrix in Sec. \ref{app:symmetrization}, and solve the only unknown parameter $\kappa$ in Sec. \ref{app:kappa}. 
Furthermore, to explain the nonlinear variation of the SKR within the range of $\kappa$, we introduce an attack model in Sec. \ref{app:model} and provide a physical interpretation. 
Finally, in Sec. \ref{app:Simulation} we provide a series of simulations, showing that the schemes using high-order constellations can achieve performance close to that of the ideal Gaussian protocol without increasing computational complexity. These results demonstrate the advantages in both high performance and efficiency, suggesting the potential for large-scale applications of DM CV-QKD systems.

\subsection{$N$-mode entangled source}\label{app:Nmode}

\begin{figure}
    \centering
    \includegraphics[width= 6 cm]{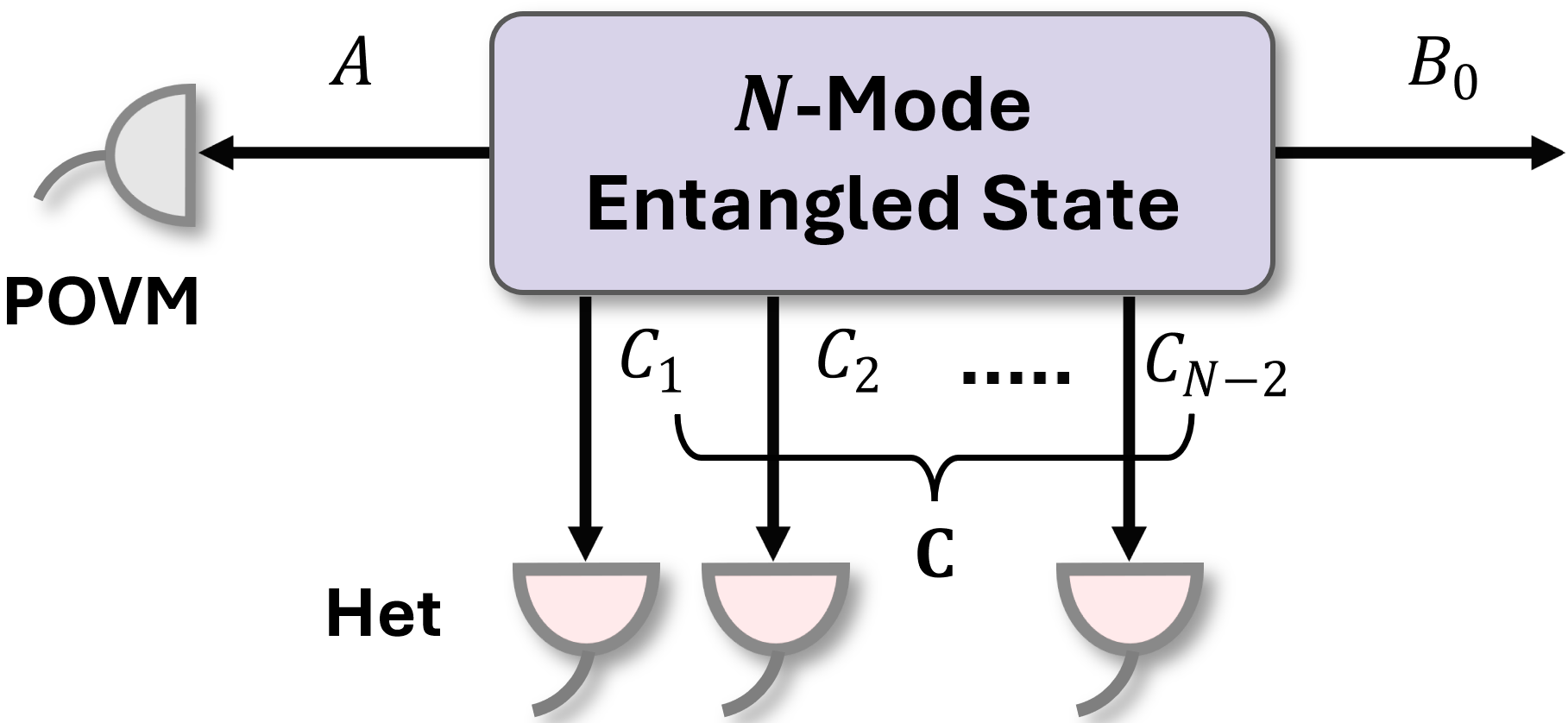}
    \caption{The $N$-mode entangled source. The source generates an \(N\)-mode entangled state, where mode \(A\) is measured by positive-operator valued measure (POVM), and modes \(\mathbf{C}=C_1C_2\dots C_{N-2}\ (N > 2)\) are measured by heterodyne detection (Het). mode \(B_0\) is sent to the receiver side.
    }\label{app:NmodeEB}
\end{figure}

Consider the EB scheme of a protocol, the state sent by the transmitter Alice usually depends on the entangled source she prepares and the measurements she performs. The most common case is Alice performing heterodyne detection on one mode of an ideal two-mode squeezed vacuum state, thereby projecting the other mode onto an infinite Gaussian state.
While to send a finite discrete-distributed coherent state, POVM measurement other than heterodyne detection should be introduced at the source. However, POVM measurement alone is insufficient to complete the security analysis. So we choose to inject more entangled modes $\mathbf{C}$ at the source to compete with eavesdropper and assist the security analysis. 
The $N$-mode entangled source is shown in Fig. \ref{app:NmodeEB}. Alice generates an $N$-mode entangled state $|\Psi\rangle_{A\mathbf{C}B_0}$, which is a purification of the mixed state $\rho_{B_0}$, where \(\mathbf{C}=C_1C_2\dots C_{N-2}\ (N > 2)\). Alice keeps modes $A$ and $\mathbf{C}$, and sends mode $B_0$ to the receiver Bob. The measurement for mode $A$ is POVM, obtaining the result $a$, and the measurements for modes $C$ are heterodyne measurements, with results recorded as $\mathbf{c}=\{c_1,\ c_2,\ \dots,\ c_{N-2}\}$. These measurements will project mode $B_0$ onto a state $\rho_{B_0}^{a,\mathbf{c}}=\{\rho_{B_0}^1,\ \rho_{B_0}^2,\ \dots,\ \rho_{B_0}^M,\ \}$.
\subsection{The sub-state $\rho_{\mathbf{C}B_0}^a$ is a product state}\label{app:POVM}
Since the sending state $\rho_{B_0}$ is only determined by the POVM result $a$ and is independent of the heterodyne detection results $\mathbf{c}$, it is necessary to require that the sub-state $\rho_{\mathbf{C}B_0}^a$ conditioned on $a$ should be a product state, that is $\rho_{\mathbf{C}B_0}^a=\rho_{\mathbf{C}}^a\otimes\rho_{B_0}^a$.
To explain this necessity, we first prove a lemma which is

\begin{lemma}
For any two-mode entangled state $\rho_{AB}$, if perform heterodyne detection on mode $A$, mode $B$ is projected onto coherent states, then the number of possibly projected coherent states for mode $B$ is either one or infinite.
\end{lemma}

\begin{proof}
We prove this by contradiction. Suppose for $\rho_{AB}$, the heterodyne detection on $A$ can project $B$ onto a finite set of coherent states
\(\{|\alpha_B^1\rangle,\dots,|\alpha_B^M\rangle\}\) with $\infty>M\ge 2$. For generality assume $\rho_{AB}$ a mixed state and exists a purification $|\Psi_{FAB}\rangle=\sum_{i=1}^M \sqrt{p_i}\,|\varphi_{FA}^i\rangle\otimes|\phi_B^i\rangle$. The heterodyne detection projects $A$ onto a coherent state $|\alpha_A\rangle$. 
We divide the entire phase space of mode $A$ into $M+1$ different non-overlapping sets \(\{\mathcal{R}_1,\dots,\mathcal{R}_M,\mathcal{R}_{M+1}\}\). The first \(M\) sets correspond to the \(M\) different output coherent states of mode \(B\), which means for a heterodyne measurement result \(\alpha_A\), if \(\alpha_A\in\mathcal{R}_i\) then mode \(B\) is projected onto \(|\alpha_B^i\rangle\). We assume that for every point in the first \(M\) sets the probability of obtaining the corresponding heterodyne measurement result is nonzero. The final set \(\mathcal{R}_{M+1}\) contains points that never occur as heterodyne results, meaning \(\operatorname{Tr}_B\bigl[\langle\alpha_A|\rho_{AB}|\alpha_A\rangle\bigr]=0\) whenever \(\alpha_A\in\mathcal{R}_{M+1}\). 
For $k\in\{1,\dots,M\}$ and $\alpha_A\in\mathcal{R}_k$, we can obtain
\begin{equation}\label{app:eq:decomp}
|\alpha_B^k\rangle\langle\alpha_B^k|
=\sum_{i=1}^M\sum_{j=1}^M C_{ij}\,|\alpha_B^i\rangle\langle\alpha_B^j|,
\end{equation}
where $C_{ij}=\sqrt{p_i p_j}/p(\alpha_A)\cdot \operatorname{Tr}_F\bigl[\langle\alpha_A|\varphi_{FA}^i\rangle\langle\varphi_{FA}^j|\alpha_A\rangle\bigr]$, $p(\alpha_A)>0$ is the probability of obtaining result $|\alpha_A\rangle$.

First, we prove that if $i\neq k$ or $j\neq k$, then $C_{ij}=0$, and the only non-zero term is $C_{kk}=1$. The intuitive understanding of this is that there are infinite equations constraining finite variables $C_{ij}$. Apply the displacement operator $D(-\alpha_B^k)$ to $|\alpha_B^k\rangle\langle\alpha_B^k|$, and isolate the vacuum term,
\begin{equation}\label{app:eq:13}
(1-C_{kk})|0\rangle\langle 0|
= \sum_{\substack{ij=11\\ ij\neq kk}}^{MM} C_{ij}\,|\alpha_B^i-\alpha_B^k\rangle\langle\alpha_B^j-\alpha_B^k|.
\end{equation}
Then calculate the inner product between the Fock state $|t\rangle$ and the vacuum state in Eq. \eqref{app:eq:13}, we have
\begin{equation}\label{app:eq:14}
0=\sum_{\substack{ij=11\\ ij\neq kk}}^{MM} C_{ij}\,
e^{-|\beta_i|^2\big/2}\,e^{-|\beta_j|^2\big/2}\,\frac{\beta_i^t\beta_j^t}{t!},\quad \forall t\ge 1,
\end{equation}
where $\beta_i=\alpha_B^i-\alpha_B^k$ and $ \beta_j=\alpha_B^j-\alpha_B^k$.
If we set $\lambda_{ij}=\beta_i\beta_j^*$ and $d_{ij}=C_{ij}\exp\bigl[-(|\beta_i|^2+|\beta_j|^2)/2\bigr]$, then the first $M^2-1$ equations can be written in the matrix formula $\Lambda \mathbf{D}= 0$, where $\Lambda$ is the transpose of a Vandermonde-like matrix with $M^2-1$ different non-zero $\lambda_{ij}$, and $\mathbf{D}$ is a vector with $M^2-1$ different $d_{ij}$. Since the Vandermonde matrix has the feature of full rank, so the only solution of \(\Lambda\mathbf{D}=0\) is \(d_{ij}=0\), which implies $\forall ij\neq kk,\ C_{ij}=0$, and $C_{kk}=1$.

Second, for any $\mathcal{R}_k$ ($1\le k\le M+1$), we can define a corresponding set $\mathcal{T}_k$ of coherent states, which is $\mathcal{T}_k=\{|\alpha_A\rangle\;|\;\alpha_A\in \mathcal{R}_k\}$. Then the above discussion leads to the conclusion that, for each $\mathcal{T}_k$, there exists at least one state orthogonal to it. First, at the case $1\le k\le M$, in which $C_{ii}=0$ for any $i=j\neq k$. This means $\forall |\alpha_A\rangle\in\mathcal{T}_k$, $\langle\alpha_A|\rho_A^{ii}|\alpha_A\rangle=0$, where $\rho_A^{ii}=\operatorname{Tr}_F\bigl[|\varphi_{FA}^i\rangle\langle\varphi_{FA}^i|\bigr]$. Then $\rho_A^{ii}$ is orthogonal to the set $\mathcal{T}_k$. Second, for the case $k=M+1$, from its definition we know $\operatorname{Tr}_B\bigl[\langle\alpha_A|\rho_{AB}|\alpha_A\rangle\bigr]=0$ if $|\alpha_A\rangle\in \mathcal{R}_{M+1}$. Then $\rho_A=\operatorname{Tr}_B[\rho_{AB}]$ is orthogonal to $\mathcal{T}_{M+1}$.

However, it can be proved that when $M$ is finite, among all these $M+1$ sets $\{\mathcal{T}_1,\mathcal{T}_2,\dots,\mathcal{T}_{M+1}\}$, at least one of them is a complete or over-complete set of the Fock state space, which means no state can be orthogonal to this set. This is contradictory to the previous conclusion. Therefore, for the case $\infty>M\ge 2$, no such a two-mode entangled state can be found. 
\end{proof}

This lemma can be easily generalized to the $N$-mode entangled state case: for any $N$-mode entangled state $\rho_{\mathbf{A}B}$, if after the heterodyne detections for each mode $\mathbf{A}$, mode $B$ is projected onto a coherent state, then the number of possibly projected coherent states for mode $B$ is either one or infinite.

For our EB scheme, the sub-state $\rho_{\mathbf{C}B_0}^a$ after the POVM measurement on mode $A$ faces the same situation as the above argument. Since which coherent state will be sent to Bob is decided by the results of POVM, then the number of possibly projected coherent states for mode $B_0$ after the heterodyne detection on mode $\mathbf{C}$ is only one. This means the conditioned sub-state $\rho_{CB_0}^{a}$ will be a product state, that is $\rho_{CB_0}^{a}=\rho_{C}^{a}\otimes |\alpha_{B_0}^{a}\rangle\langle\alpha_{B_0}^{a}|$.  

The above conclusion illuminates the  logical structure of our framework: First, POVM measurement should be introduced to prepare finite coherent states. Consequently, entangled state with more than
two modes are necessary to assist security analysis. Furthermore, heterodyne detection performed on those auxiliary modes does not change the sending coherent state, which is determined only by the POVM results. Therefore, the conditioned sub-state after the POVM measurement should be a product state.


\subsection{The simplified representation of $\rho_C$}\label{app:simply_rhoC}

The secret key rate calculation is based on the covariance matrix $\gamma_{ACB}$, in which $\gamma_A$, $\gamma_C$, and $\psi_{AC}$ can be theoretically calculated, and $\gamma_B$ is estimated only using Bob's data. Only the estimation of $\psi_{CB}$ will use the measurement results of mode $C$. Generally, the measurement result $c$ should be another set of random numbers in the PM model, which can be simulated through a quantum random number generator (QRNG). However, the product feature of sub-state $\rho_{CB_0}^a$ can help to simplify this.

Let's take the x-quadrature of mode $C$ as an example. The state of modes $C$ and $B_0$ is $\rho_{C B_0} = \sum_{i=1}^M p_i\, \rho_{C}^i \otimes \rho_{B_0}^i$.
Using the product structure of the sub-states, the covariance between $\hat{x}_{C}$ and $\hat{x}_{B_0}$ can be written as
\begin{equation}
\label{app:eq:covariance_trace}
\begin{aligned}
    &\bigl\langle \Delta \hat{x}_{C}\,\Delta \hat{x}_{B_0}\bigr\rangle \\
&= \sum_{i=1}^M p_i\mathrm{Tr}_{C B_0}\Bigl[ \bigl(\hat{x}_{C}-\bar{x}_{C}\bigr)\bigl(\hat{x}_{B_0}-\bar{x}_{B_0}\bigr)\,\rho_{C}^i\otimes\rho_{B_0}^i \Bigr] \\
&=\sum_{i=1}^M p_i\,
\bigl(\bar{x}_{C}^i-\bar{x}_{C}\bigr)\bigl(\bar{x}_{B_0}^i-\bar{x}_{B_0}\bigr),
\end{aligned}
\end{equation}
where $\bar{x}_{C}^i = \mathrm{Tr}_{C}\bigl(\hat{x}_{C}\,\rho_{C}^i\bigr)$ and $\bar{x}_{C} = \sum_{i=1}^M p_i\,\bar{x}_{C}^i$ can be theoretically calculated. Therefore, if $\rho_{B_0}$ has been simulated by a sequence of random number, then it's enough to let the other sequence be $\{\bar{x}_{C}^i,\ \bar{p}_{C}^i\}$ for the calculation of $\psi_{CB}$. This is much simpler than simulating the heterodyne results $c$ using quantum random numbers.

\begin{figure}
    \centering
    \includegraphics[width= 8.5 cm]{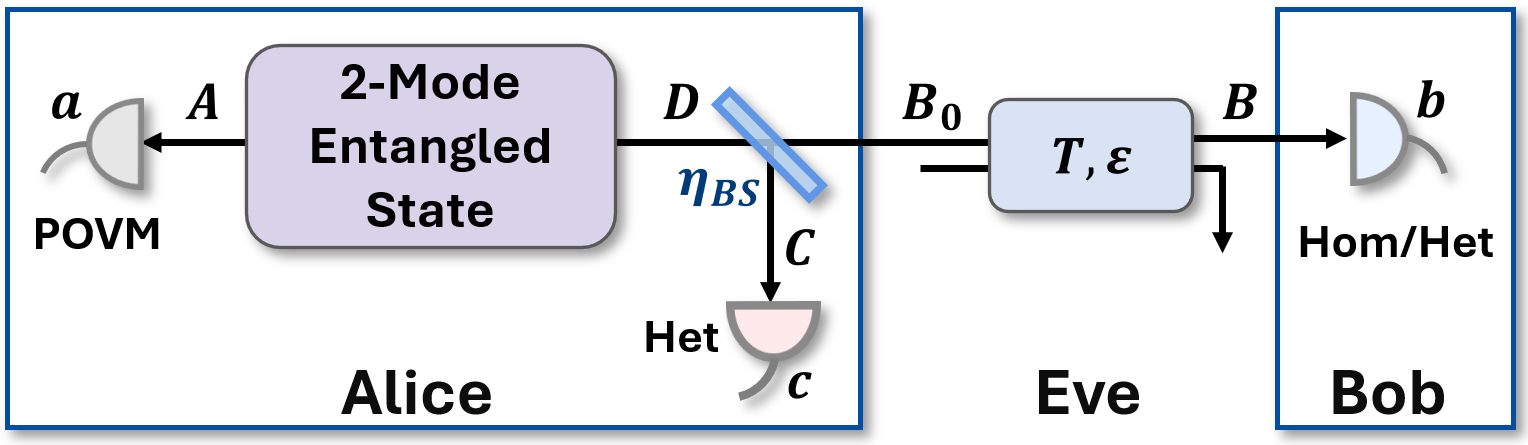}
    \caption{The Entanglement-based (EB) scheme of the proposed DM CV-QKD protocol with a three-mode entangled source. POVM, positive-operator valued measurement; Het, heterodyne measure. Alice prepares a two-mode entangled state $|\Psi_{AD}\rangle$ and performs a POVM measurement on mode $A$, mode $D$ is split by a beam splitter with transmittance $\eta_{BS}$ into modes $C$ and $B_0$. Alice performs heterodyne detection on mode $C$. Mode $B_0$ is transmitted through a quantum channel fully controlled by the eavesdropper Eve with transmittance $T$ and excess noise $\varepsilon$, and heterodyne detected by Bob. Here $a$, $c$ and $b$ are measurement results for $A$, $C$ and $B$ respectively.
    }\label{app:EB}
\end{figure}

\subsection{Three-mode discrete-modulated source}\label{app:SourceDesign}

In this subsection, we detail the design principles for an equivalent entanglement source compatible with any discrete-modulation constellation. A three-mode entangled source is adopted as shown in Fig. \ref{app:EB}, which is not only structurally simple but also highly effective.

\subsubsection{Design of two-mode entangled state}
Our design principle is to let the correlation between modes $C$ and $B_0$ as strong as possible, which from the covariance matrix perspective we want \(\langle\Delta\hat{x}_C\Delta\hat{x}_{B_0}\rangle\) and \(\langle\Delta\hat{p}_C\Delta\hat{p}_{B_0}\rangle\) to be as large as possible.
Suppose the source consists of \(M\) distinct coherent states $S_{\rho}=\bigl\{\,|\alpha_{B_0}^1\rangle,\;|\alpha_{B_0}^2\rangle,\;\dots,\;|\alpha_{B_0}^M\rangle\,\bigr\}$, and the three-mode entangled source is denoted by \(|\Psi_{ACB_0}\rangle\).
Assume the covariance matrix \(\gamma_{ACB_0}\) of \(|\Psi_{ACB_0}\rangle\) is in the standard form (so that \(\langle\Delta\hat{x}_i\Delta\hat{p}_j\rangle=0\) for all \(i,j\)), which can be arranged in the QAM case.

We take the \(x\)-quadrature as an example. Denote the mean and variance each substate \(\rho_C^i\) and of the coherent state \(|\alpha_{B_0}^i\rangle\) by
\begin{equation}\label{app:eq:15}
\begin{aligned}
\bar{x}_C^i &= \operatorname{Tr}\bigl[\hat{x}_C\,\rho_C^i\bigr], \quad
V_{C,x}^i = \operatorname{Tr}\bigl[\hat{x}_C^2\,\rho_C^i\bigr] - (\bar{x}_C^i)^2,\\
\bar{x}_{B_0}^i &= \langle\alpha_{B_0}^i|\hat{x}_{B_0}|\alpha_{B_0}^i\rangle, \quad
V_{B_0,x}^i = 1.
\end{aligned}
\end{equation}
The overall mean values for modes \(C\) and \(B_0\) are $\bar{x}_C=\sum_{i=1}^M p_i\,\bar{x}_C^i$ and $
\bar{x}_{B_0}=\sum_{i=1}^M p_i\,\bar{x}_{B_0}^i$, 
and the corresponding variances are
\begin{equation}\label{app:eq:16}
\begin{aligned}
V_{C,x} &= \sum_{i=1}^M p_i\,V_{C,x}^i \;+\; \sum_{i=1}^M p_i\bigl(\bar{x}_C^i-\bar{x}_C\bigr)^2,\\
V_{B_0,x} &= 1 \;+\; \sum_{i=1}^M p_i\bigl(\bar{x}_{B_0}^i-\bar{x}_{B_0}\bigr)^2.
\end{aligned}
\end{equation}
Given that the sub-state $\rho_{CB_0}^a$ is a product state, the covariance between \(C\) and \(B_0\) in the \(x\)-quadrature is
\begin{equation}\label{app:eq:17}
\begin{aligned}
&\bigl|\langle\Delta\hat{x}_C\Delta\hat{x}_{B_0}\rangle\bigr|= \bigl|\langle(\hat{x}_C-\bar{x}_C)(\hat{x}_{B_0}-\bar{x}_{B_0})\rangle\bigr|\\
&= \Biggl|\sum_{i=1}^M p_i\,
\operatorname{Tr}_{CB_0}\bigl[(\hat{x}_C-\bar{x}_C)(\hat{x}_{B_0}-\bar{x}_{B_0})
\,\rho_C^i\otimes\rho_{B_0}^i\bigr]\Biggr|\\
&= \Biggl|\sum_{i=1}^M p_i\bigl(\bar{x}_C^i-\bar{x}_C\bigr)\bigl(\bar{x}_{B_0}^i-\bar{x}_{B_0}\bigr)\Biggr|\\
&\le \sqrt{\Bigl(\sum_{i=1}^M p_i\bigl(\bar{x}_C^i-\bar{x}_C\bigr)^2\Bigr)
\Bigl(\sum_{i=1}^M p_i\bigl(\bar{x}_{B_0}^i-\bar{x}_{B_0}\bigr)^2\Bigr)}\\
&= \sqrt{\bigl(V_{C,x}-\sum_{i=1}^M p_i V_{i,Cx}\bigr)\bigl(V_{B_0,x}-1\bigr)}.
\end{aligned}
\end{equation}
The inequality above follows from the Cauchy-Schwarz inequality. Equality holds if and only if there exists a nonzero scalar \(t\) such that for all \(i\), $(\bar{x}_C^i-\bar{x}_C)/(\bar{x}_{B_0}^i-\bar{x}_{B_0}) \equiv t$.
By the uncertainty relation each sub-state satisfies \(V_{C,x}^i V_{C,p}^i\ge 1\). If we further assume \(x\) and \(p\) are symmetric for each substate (so \(V_{C,x}^i\ge 1\) and \(\sum_i p_i V_{C,x}^i\ge 1\)), then maximizing \(|\langle\Delta\hat{x}_C\Delta\hat{x}_{B_0}\rangle|\) requires \(V_{C,x}^i=1\) for all \(i\). These conditions are simultaneously met when each \(\rho_C^i\) is itself a coherent state \(|\alpha_C^i\rangle\) and the means satisfy a linear relation $\alpha_C^i/\alpha_{B_0}^i = t,\ \forall i\in\{1,\dots,M\}$.

Such a linear relationship can be realized by a beam splitter model: prepare a coherent state \(|\beta_D^i\rangle\) with $\beta_D^i=\sqrt{1+t^2}\,\alpha_{B_0}^i$, and send it through a beam splitter of transmittance \(\eta_{BS}=1/(1+t^2)\). In this construction Alice only needs to prepare a two‑mode entangled state of the form $|\psi_{AD}\rangle=\sum_{i=1}^M \sqrt{p_i}\,|R_A^i\rangle\,|\beta_D^i\rangle$, and then let mode \(D\) pass through the beam-splitter to produce the desired correlations between modes \(C\) and \(B_0\), as illustrated in Fig. \ref{app:EB}.

For the explicit design of \(|\psi_{AD}\rangle\), first symmetrize the mixed state
\(\rho_D=\sum_{i=1}^M p_i\,|\beta_D^i\rangle\langle\beta_D^i|\) in an orthonormal basis \(\{|\theta_D^i\rangle\}\):
\begin{equation}\label{app:eq:18}
\rho_D=\sum_{i=1}^M p_i\,|\beta_D^i\rangle\langle\beta_D^i|
=\sum_{i=1}^M v_i\,|\theta_D^i\rangle\langle\theta_D^i|.
\end{equation}
Write the orthonormal vectors as $|\theta_D^i\rangle=\sum_{j=0}^{\infty} c_{ij}\,|j\rangle$, and define
\begin{equation}\label{app:eq:19}
|\psi_{AD}\rangle=\sum_{i=1}^M \sqrt{v_i}\,|\varphi_A^i\rangle\,|\theta_D^i\rangle,
\end{equation}
where the states \(|\varphi_A^i\rangle\) are designed as $|\varphi_A^i\rangle=\sum_{j=0}^{\infty} (c_{ij})^*\,|j\rangle$.


\begin{figure*}[ht]
\centering
\includegraphics[width=14cm]{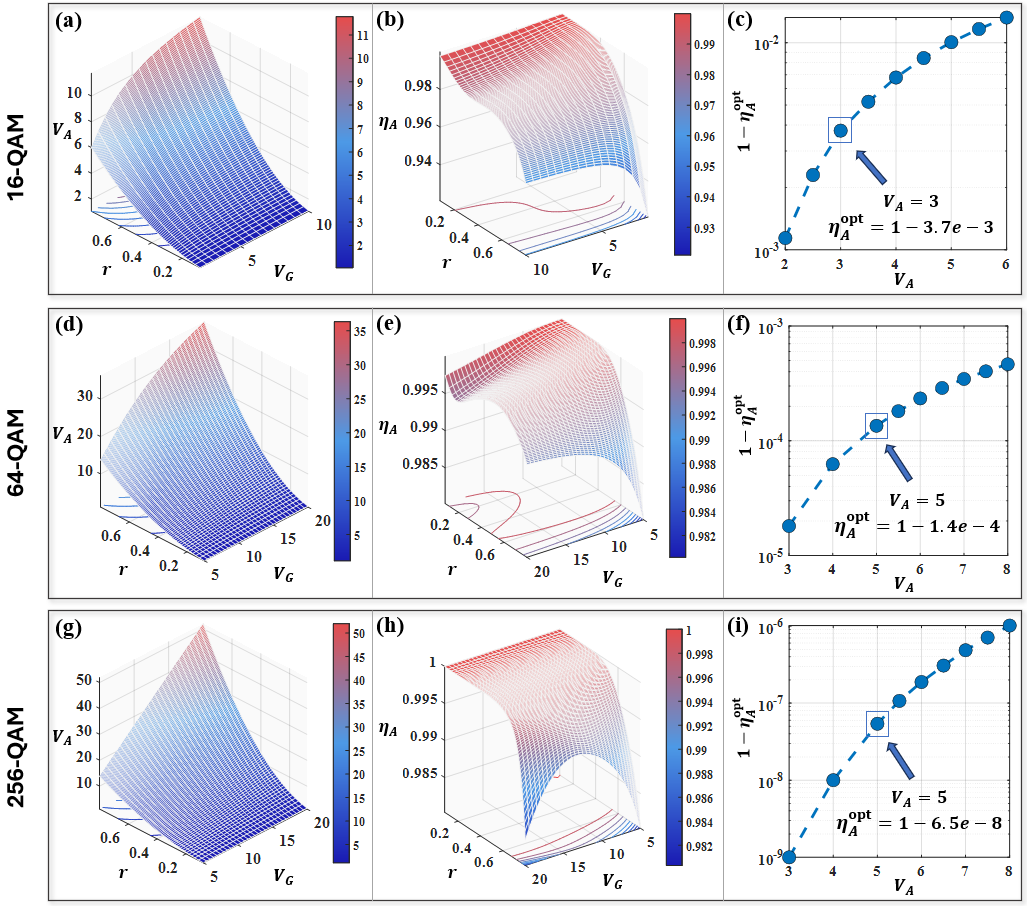}
\caption{Performance of the entangled source with 16-, 64-, 256-QAM constellation using Gaussian probability constellation shaping.}
\label{app:source_all}
\end{figure*}

\subsubsection{Optimization for quadrature-amplitude modulation cases}
QAM format is a standard modulation format in classical coherent communications and has been widely applied. Systems employing this modulation are naturally compatible with the existing electro-optical device.

For a standard $M$-QAM ($M=L^2$, with $L$ a positive integer) format, coherent states are placed at the intersections of \(L\) equally spaced columns and \(L\) equally spaced rows in phase space (or classically called constellation map). If the space between adjacent columns (or rows) is \(2r\), then the positions of the \(M\) coherent states can be written as
\begin{equation} 
    \{\forall \mu, \nu \in [1,L], \alpha_{\mu \nu}=(2\mu-1-L)r+i \cdot (2\nu-1-L)r\}. 
\end{equation}
For this standard QAM constellation, the covariance matrix of \(|\Psi_{AD}\rangle\) exhibits the following form,
\begin{equation} 
    \gamma_{AD}=
    \begin{bmatrix} 
            V_AI_2 & \phi_{AD}\sigma_Z \\ 
    \phi_{AD}\sigma_Z & V_DI_2 
    \end{bmatrix}, 
\end{equation}
where $I_2=
\begin{bmatrix}
		1 & 0 \\
		0 & 1
\end{bmatrix}$, $\sigma_Z=
\begin{bmatrix}
		1 & 0 \\
		0 & -1
\end{bmatrix}$. According to the uncertainty principle, we know $\phi_{AD} \le \sqrt{V_A^2-1}$, then an ideality coefficient $\eta_A=\phi_{AD}^2\Big/(V_A^2-1)$ can be defined to quantify the deviation of $|\Psi_{AD}\rangle$ from a two-mode squeezed state (TMSV). This deviation depends on three factors: the number of coherent states $M$, their positions in phase space $\alpha_{\mu \nu}$, and their probability distribution $p(\alpha_{\mu \nu})$.

Using Gaussian probability constellation shaping to optimize the distribution of these states, yields that
\begin{equation} 
    p(\alpha_{\mu \nu})=e^{\frac{-\vert\alpha_{\mu \nu}\vert^2}{(2V_G)}}\Big/\sum{e^{\frac{-|\alpha_{\mu \nu}|^2}{(2V_G)}}},
\end{equation}
where $V_G$ is a parameter related to probability shaping. 

The above processes specify the quantum state $|\Psi_{AD}\rangle$ to be transmitted.
We numerically calculate the \(\eta_A\) and $V_A$ for 16-QAM, 64-QAM and 256-QAM with different \(V_G\) and \(r\), to find a relatively optimal combination of \(V_A\) and \(\eta_A\), as shown in  Fig. \ref{app:source_all}. 
We first use a heatmap to illustrate the relationship between $V_A$ and \(V_G\) and \(r\) in Fig. \ref{app:source_all} (a) (d) and (g). Then we calculate the corresponding $\eta_A$ for given values of $V_G$ and $r$ in Fig. \ref{app:source_all} (b) (e) and (h), which allows us to determine the optimal $\eta_A^{\mathrm{opt}}$ for any given $V_A$, and we display the relationship of $V_A$ and $1-\eta_A^{\mathrm{opt}}$ in Fig. \ref{app:source_all} (c) (f) and (i). 
It can be seen that the more coherent states used and the smaller the modulation variance is set, the closer $|\Psi_{AD}\rangle$ approaches the ideal TMSV state. Considering both the performance degradation caused by the nonideal discrete-modulated source and the signal-to-noise ratio requirements of practical experiments, the typical $(V_G, V_A, \eta_A^{\mathrm{opt}})$ for three modulation formats can be selected as shown in Tab. \ref{app:tab1}.

\begin{table}[ht]
\centering
\caption{Typical $V_A$ and corresponding optimal $\eta_A^{\mathrm{opt}}$.}
\label{app:tab1}
\begin{tabular}{@{} l c c c @{}}
\toprule
\textbf{Constellation} & \(\mathbf{V_G}\) & \(\mathbf{V_A}\) & \(\boldsymbol{\eta_A^{\mathrm{opt}}}\) \\
\midrule
16-QAM & 3  & 3 & \(1-3.7\times10^{-3}\) \\
64-QAM & 6  & 5 & \(1-1.4\times10^{-4}\) \\
256-QAM & 11 & 5 & \(1-6.5\times10^{-8}\) \\
\bottomrule
\end{tabular}
\end{table}

It is worth mentioning that the quantum state sent into the channel is determined by the design of $|\psi_{AD}\rangle$, which implies that the protocol can be customized following a specific system — one may first identify the quantum state required in practice and then optimize the corresponding two-mode entangled source. Taking the two-mode squeezed vacuum (TMSV) as an ideal reference, the gap between \(|\Psi\rangle_{AD}\) and TMSV constitutes the primary factor influencing protocol performance. Therefore, from a practical engineering perspective, we concentrate on the design of standard quadrature-amplitude-modulation (QAM) modulation source used in classical coherent communications, since systems employing this modulation are particularly well aligned with the existing electro-optic devices.

\subsection{Secret key rate calculation based on covariance matrix}\label{app:SKR}
Here we explain the derivation of the secret key rate formula for our three-mode protocol in Fig. \ref{app:EB}. It should be noted that the method is not limited to the three-mode case and can be generalized to the cases where Alice has an arbitrary $N\ (\ge 2)$ mode entangled source.

A generally used secret key rate for the asymptotic case is the Devetak--Winter formula \cite{app:devetak2005distillation},
\begin{equation}\label{app:eq:DW}\tag{3}
K=\beta I(a:b)-S(b:E),
\end{equation}
where \(I(a:b)\) is the classical mutual information between Alice and Bob, \(\beta\) is the classical reconciliation efficiency, and \(S(b:E)\) is the Holevo information \cite{app:holevo1973bounds} between Bob's data and the adversary Eve. Usually, \(S(b:E)\) can be replaced by any of its upper bounds, among which the upper bound \(S_{BE}^G\) derived from the Gaussian state extremality theorem \cite{app:garcia2006unconditional,app:wolf2006extremality}  is the most commonly used case, because its calculation only relies on the covariance matrix \(\gamma_{ACB}\), which can be estimated through the experimental data.

The covariance matrix \(\gamma\) of an \(N\)-mode state \(\hat{\rho}_N\) is defined as \cite{app:weedbrook2012gaussian}
\begin{equation}\label{app:eq:cov_def}\tag{4}
\gamma_{ij}:=\tfrac{1}{2}\big\langle\{\Delta\hat{r}_i,\Delta\hat{r}_j\}\big\rangle,
\end{equation}
where \(\hat{r}=\{\hat{x}_1,\hat{p}_1,\ldots,\hat{x}_N,\hat{p}_N\}\), \(\langle\hat{r}_i\rangle=\mathrm{Tr}(\hat{r}_i\hat{\rho}_N)\), and \(\Delta\hat{r}_i=\hat{r}_i-\langle\hat{r}_i\rangle\). 
Suppose \(\gamma_{ACB}\) is the covariance matrix of the state \(\rho_{ACB}\), which is the state after mode \(B_0\) of the entangled source \(\ket{\Psi_{ACB_0}}\) arriving at Bob's side through the channel. It can be represented using several sub-matrices,
\begin{equation}\label{app:eq:gamma_ACB}\tag{5}
\gamma_{ACB}=
\begin{pmatrix}
\gamma_A & \psi_{AC} & \kappa_{AB} \\
\psi_{AC}^T & \gamma_C & \psi_{CB} \\
\kappa_{AB}^T & \psi_{CB}^T & \gamma_B
\end{pmatrix}.
\end{equation}
Here \(\gamma_A,\gamma_C\) and \(\psi_{AC}\) can be theoretically calculated from \(\Psi_{ACB_0}\), since modes \(A\) and \(C\) are kept on Alice's side. \(\psi_{CB}\) and \(\gamma_B\) can be estimated after Alice and Bob randomly share part of their coherent measurement results. The only unknown sub-matrix is \(\kappa_{AB}\), since the measurements for modes \(A\) are not coherent measurements now.

Nevertheless, the uncertainty principle \cite{app:weedbrook2012gaussian} imposes a constraint on  the covariance matrix \(\gamma_N\) for an \(N\)-mode state, which is
\begin{equation}\label{app:eq:uncertainty}\tag{6}
\gamma_N + i\Omega_N \ge 0,
\end{equation}
where \(\Omega_N=\mathrm{diag}(\omega_1,\omega_2,\ldots,\omega_N)\), and
\begin{equation}\label{app:eq:omega}\tag{7}
\omega_1=\cdots=\omega_N=\omega=
\begin{pmatrix}
0 & 1\\
-1 & 0
\end{pmatrix}.
\end{equation}

We denote \(\mathcal{F}_{\kappa_{AB}}\) as the set of all \(\kappa_{AB}\) satisfying this constraint for \(\gamma_{ACB}\), that is
\begin{equation}\label{app:eq:S_k}\tag{8}
\mathcal{F}_{\kappa_{AB}}=\big\{\phi_{AB}\ \big|\ \gamma_{ACB}[\kappa_{AB}=\phi_{AB}]+i\Omega_3\ge 0\big\}.
\end{equation}
If \(\phi_{AB}^{\mathrm{worst}}\) is the real eavesdropping induced \(\kappa_{AB}\), then \(\phi_{AB}^{\mathrm{worst}}\in \mathcal{F}_{\kappa_{AB}}\). It can be understood that \(S_{BE}^G\) is a function of \(\kappa_{AB}\) now. Then by traversing the set \(\mathcal{F}_{\kappa_{AB}}\) for all possible \(\kappa_{AB}\), we can find the maximum of \(S_{BE}^G\). The secret key rate can be written as
\begin{equation}\label{app:eq:key_rate_sup}\tag{9}
K=\beta I(a:b)-\sup_{\kappa_{AB}\in \mathcal{F}_{\kappa_{AB}}} S_{BE}^G.
\end{equation}

Now we introduce the calculation method for each term. The reconciliation efficiency $\beta$ is determined by error correction codes and the signal-to-noise ratio of the measurement results. For \(\beta I(a:b)\), it can be expressed as \(\beta I(a:b)=H(b)-\mathrm{leak}_{\mathrm{EC}}\), in which \(H(b)\) is the Shannon entropy of Bob's measurement results, and \(\mathrm{leak}_{\mathrm{EC}}\) represents the information that Bob sends to Alice for the data reconciliation. Both these terms can be obtained from measured data and the error correction step. Given the channel model, the probability \(p(b\mid i)\) can be obtained from the model, which is the probability of getting Bob's measurement result \(b\) given Alice sending the state \(\rho_B^i\). The overall probability is \(p(b)=\sum_{i=1}^M p_i\,p(b\mid i)\). Since \(b\) is a discrete variable quantized from the measurement result, then
\begin{equation}\label{app:eq:mutual_discrete}\tag{10}
\begin{aligned}
I(a:b)=&-\sum_{b} p(b)\log_2 p(b)\\
&+\sum_{i=1}^M p_i \sum_{b} p(b\mid i)\log_2 p(b\mid i).
\end{aligned}
\end{equation}

For \(\sup_{\kappa_{AB}\in \mathcal{F}_{\kappa_{AB}}} S_{BE}^G(\kappa_{AB})\), we traverse each \(\mathcal{F}_{\kappa_{AB}}\) to calculate its corresponding \(S_{BE}^G(\kappa_{AB})\) and find the maximal value. \(S_{BE}^G(\kappa_{AB})\) can be expressed as $S_{BE}^G(\kappa_{AB})= S\bigl(\rho_{A B C}^G\bigr)-S\bigl(\rho_{A C\mid b}^G\bigr)$,
where \(\rho_{A C B}^G\) is the Gaussian state with the same covariance matrix as \(\gamma_{ACB}(\kappa_{AB})\), and \(\rho_{A C\mid b}^G\) denotes the conditional Gaussian state (with covariance \(\gamma_{AC\mid b}\)) related to Bob's measurement method. The procedures to obtain \(\gamma_{AC\mid b}\) from \(\gamma_{ACB}\) and to evaluate von Neumann entropies for Gaussian states can be found in \cite{app:weedbrook2012gaussian,app:fossier2009improvement}.

\begin{figure}
    \includegraphics[width= 8.5 cm]{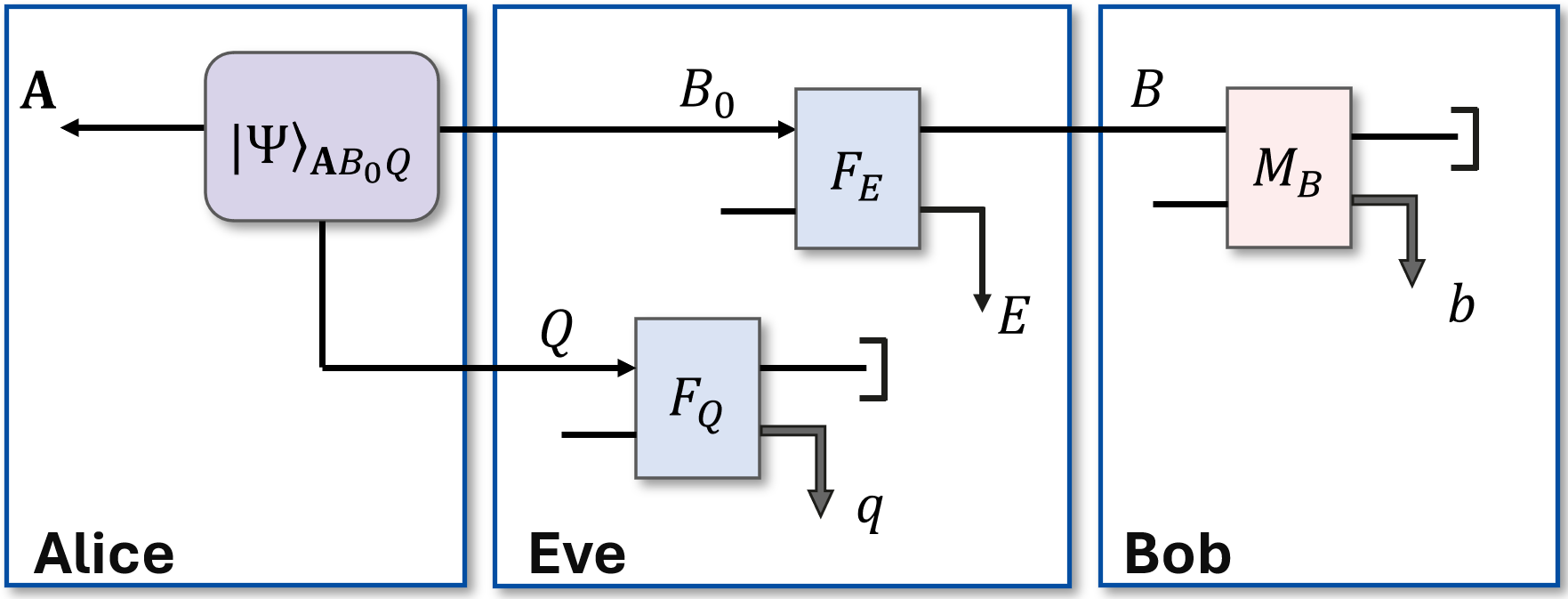}
    \caption{Mixture and measurement of two entangled sources.
    }\label{app:Mix}
\end{figure}

\subsection{Convexity of secret key rate}\label{app:Convexity}
Here we describe the convexity of SKR as the following lemma.
\begin{lemma}
Let $\rho_{\mathbf{A}B}^{(+)}$ and $\rho_{\mathbf{A}B}^{(-)}$ be two $N$-mode quantum states shared by Alice and Bob, where $\mathbf{A}=A_1A_2\cdots A_{N-1}$ denotes the modes held by Alice and $B$ denotes Bob's mode. Assume that the two states yield the same SKR,
\begin{equation}
K\left(\rho_{\mathbf{A}B}^{(+)}\right)=
K\left(\rho_{\mathbf{A}B}^{(-)}\right)
=K_0.
\end{equation}
For any probabilistic mixture
\begin{equation}
\rho_{\mathbf{A}B}=
p_1\rho_{\mathbf{A}B}^{(+)}+
p_2\rho_{\mathbf{A}B}^{(-)},
\quad
p_1+p_2=1,
\end{equation}
the SKR of the mixed state satisfies
\begin{equation}
K(\rho_{\mathbf{A}B})
\le p_1K\left(\rho_{\mathbf{A}B}^{(+)}\right)+
p_2K\left(\rho_{\mathbf{A}B}^{(-)}\right)=
K_0.
\end{equation}
It can be expressed as
\begin{equation} 
    K\big(\sum_{i=1}^2 p_i \rho_i\big) \le \sum_{i=1}^2 p_iK(\rho_i)=K(\rho_i),
\end{equation} 
where $\sum_i p_i=1$, $\rho_1=\rho_{\mathbf{A}B}^{(+)}$, $\rho_2=\rho_{\mathbf{A}B}^{(-)}$.
\end{lemma}

\begin{proof}
We first explain the steps for mixture of the two states, as shown in Fig. \ref{app:Mix}:

1). Suppose there are two entanglement sources \(|\Psi\rangle_{\mathbf{A}B_0}^{(+)}\) and \(|\Psi\rangle_{\mathbf{A}B_0}^{(-)}\). After the optimal Gaussian attack by the eavesdropper, they yield the target states $\rho_{\mathbf{A}B}^{(+)}$ and $\rho_{\mathbf{A}B}^{(-)}$. 
Then mix the two entanglement sources and get \(\rho_{\mathbf{A}B_0}=p\rho_{\mathbf{A}B_0}^{(+)}+(1-p)\rho_{\mathbf{A}B_0}^{(-)}\).

2). Introduce an auxiliary system \(Q\), which can purify the entire system and yield 
\begin{equation}
    |\Psi\rangle_{\mathbf{A}B_0Q}=\frac{1}{\sqrt{2}}\bigl(|\Psi\rangle_{\mathbf{A}B_0}^{(+)}|0\rangle_Q+|\Psi\rangle_{\mathbf{A}B_0}^{(-)}|1\rangle_Q\bigr),
\end{equation}
and this is a purification of the mixed state $\rho_{\mathbf{A}B_0}$. Then a POVM measurement $F_Q$ is performed on \(Q\), obtaining the result $q$, with the result $q=|0\rangle_Q$ projecting the subsystem \(ACB_0\) onto \(|\Psi\rangle_{\mathbf{A}B_0}^{(+)}\), and $q=|1\rangle_Q$ projecting it onto $|\Psi\rangle_{\mathbf{A}B_0}^{(-)}$.

3). The mixed state $\rho_{B_0}$ is sent to Bob  through the quantum channel. Eve implements an attack described by a unitary $F_E$ and can purify the system $\rho_{\mathbf{A}B Q}$, so that the entire state after the channel is
\begin{equation}
|\Psi\rangle_{\mathbf{A}B E Q}=F_E\bigl(|\Psi\rangle_{\mathbf{A}B_0 Q}\otimes|\Psi\rangle_{E}\bigr),
\end{equation}
and we can get the mixed state $\rho_{\mathbf{A}B}=p\rho_{\mathbf{A}B}^{(+)}+(1-p)\rho_{\mathbf{A}B}^{(-)}$.

4). Bob receives the state $\rho_B$ and performs a heterodyne measurement, obtaining the result $b$.

5). The measurement $q$ is published, allowing Alice and Bob to align which state they have sent and measured, and then perform post-processing.

Now we analyze the maximum information Eve can obtain. Suppose Eve can fully obtain the result $q$ and get the joint system $Eq$, then Eve's accessible information should be $S(b:Eq)$. From the definition of von Neumann mutual information, we have,
\begin{equation}\label{app:eq:mutual_def}
    \begin{aligned}
        S(b:E q)=&S(b)+S(E q)-S(b q E) \\
        =&S(\rho_b)+S(\rho_{E q})-S(\rho_{b q E}).
    \end{aligned}
\end{equation}
The calculation of \(S(b:E q)\) is based on the following classical-quantum state
\begin{equation}\label{app:eq:rho_bqE}
\rho_{b q E}=\iint \mathrm{d}b\,\mathrm{d}q\; p(b,q)\,|b\rangle\langle b|\otimes|q\rangle\langle q|\otimes\rho_E^{b q}.
\end{equation}
From this we can write
\begin{equation}\label{app:eq:rho_b}
    \begin{aligned}
        \rho_b=&\int \mathrm{d}b\; p(b)\,|b\rangle\langle b|,\\
\rho_{E q}=&\int \mathrm{d}q\; p(q)\,|q\rangle\langle q|\otimes\rho_E^{q},\\
\rho_{E q}^{b}=&\int \mathrm{d}q\; p(q\mid b)\,|q\rangle\langle q|\otimes\rho_E^{b q}.
    \end{aligned}
\end{equation}
Using von Neumann entropy and joint entropy definitions, we have
\begin{equation}\label{app:eq:S_b}
    \begin{aligned}
        S(b)=&S\bigl(\int \mathrm{d}b\; p(b)\,|b\rangle\langle b|\bigr)=H\bigl(p(b)\bigr),\\
S(E q)=&H\bigl(p(q)\bigr)+\int \mathrm{d}q\; p(q)\,S\bigl(\rho_E^{q}\bigr),\\
S(b q E)=&H\bigl(p(b)\bigr)+\int \mathrm{d}b\; p(b)\,S\bigl(\rho_{E q}^{b}\bigr).
    \end{aligned}
\end{equation}
The conditional entropy can be written as
\begin{equation}\label{app:eq:cond_entropy}
    \begin{aligned}
        \int \mathrm{d}b\; p(b)\,S\bigl(\rho_{E q}^{b}\bigr)=&\int \mathrm{d}b\; p(b)\,H\bigl(p(q\mid b)\bigr)+\\
&\int \mathrm{d}b\; p(b)\int \mathrm{d}q\; p(q\mid b)\,S\bigl(\rho_E^{b q}\bigr).
    \end{aligned}
\end{equation}
Therefore, \(S(b:E q)\) can be written as
\begin{equation}\label{app:eq:S_bEq_expand}
\begin{aligned}
S(b:E q)=&H\bigl(p(q)\bigr)+\int\mathrm{d}q\; p(q)\,S(\rho_E^{q})\\
&-\int\mathrm{d}b\; p(b)\,S(\rho_{E q}^{b}).
\end{aligned}
\end{equation}
Since \(A B C E Q\) is a pure system, for rank-1 measurements we have $S(\rho_E^{q})=S(\rho_{\mathbf{A}B}^{q})$ and $S(\rho_E^{b q})=S(\rho_{\mathbf{A}}^{b q})$. Moreover, since $H(p(q))=H(q)$, $H(q\mid b)=\int \mathrm{d}b\; p(b)\,H\bigl(p(q\mid b)\bigr)$, $I(b:q)=H(q)-H(q\mid b)$. Using these relations, we obtain a simplified form of Eve's knowledge
\begin{equation}\label{app:eq:S_bEq_simplified}
\begin{aligned}
S(b:E q)
&=I(b:q)+\int\mathrm{d}q\; p(q)\Bigl[S(\rho_{\mathbf{A}B}^{q})\\
&-\int\mathrm{d}b\; p(b\mid q)\,S(\rho_{\mathbf{A}}^{b q})\Bigr].
\end{aligned}
\end{equation}
The bracketed term has the same form as a classical--quantum mutual information, it can be written as
\begin{equation}\label{app:eq:cond_mutual}
S(E:b\mid q)= S(\rho_{\mathbf{A}B}^{q})-\int\mathrm{d}b\; p(b\mid q)\,S(\rho_{\mathbf{A}}^{b q}).
\end{equation}
When the measurement result $q$ is \(q=|0\rangle_Q\) we have
\begin{equation}\label{app:eq:cond0}
S(b:E\mid q=|0\rangle_Q)=S(b:E)\bigl(\rho_{\mathbf{A}B}^{(+)}\bigr),
\end{equation}
and when \(q=|1\rangle_Q\) we have
\begin{equation}\label{app:eq:cond1}
S(b:E\mid q=|1\rangle_Q)=S(b:E)\bigl(\rho_{\mathbf{A}B}^{(-)}\bigr).
\end{equation}
Since $S(E:b)\bigl(\rho_{\mathbf{A}B}^{+}\bigr)=S(E:b)\bigl(\rho_{\mathbf{A}B}^{-}\bigr)=S(E:b)_1$, and $S(E:b)_1$ is defined as Eve's information obtained from either independent quantum state. Then $S(b:E q)$ can be further simplified as
\begin{equation}\label{app:eq:S_bEq_final}
\begin{aligned}
S(b:E q)
&=I(b:q)+pS(E:b)\bigl(\rho_{\mathbf{A}B}^{((+))}\bigr)+\\
&(1-p)S(E:b)\bigl(\rho_{\mathbf{A}B}^{((-))}\bigr)\\
&=I(b:q)+S(E:b)\bigl(\rho_{\mathbf{A}B}^{(+)}\bigr)\ge S(E:b)_1.
\end{aligned}
\end{equation}
Thus, the convexity of Eve's information is proven.

Our goal is to use the covariance matrix $\gamma_{\mathbf{A}B}$ of the mixed state $\rho_{\mathbf{A}B}$ to calculate the mutual information \(S(b:E Q)_{\mathrm{G}}\) when the system $\mathbf{A}BQE$ is Gaussian, and with the quantum data-processing inequality \cite{app:schumacher1996quantum} and Gaussian optimality \cite{app:garcia2006unconditional,app:wolf2006extremality}, we have
\begin{equation}\label{app:eq:QDPI}
S(b:E Q)_{\mathrm{G}} \ge S(b:E Q)\ge S(b:E q)\ge S(E:b)_1.
\end{equation}

Additionally, the mutual information between Alice and Bob is a classical result that is not affected by quantum operations. Therefore, the relationship between SKR is 
\begin{equation}
K(\gamma_{\mathbf{A}B})=K(\rho_{\mathbf{A}B}^G) \le K(\rho_{\mathbf{A}B})\le K(\rho_{\mathbf{A}B}^{(+)}).
\end{equation}
Finally, the convexity of SKR is proved.
\end{proof}

\subsection{Covariance matrix symmetrization}\label{app:symmetrization}

In this subsection, we detail the symmetrization steps for covariance matrix, as shown in Fig. \ref{app:Fig:SymStep}. Firstly, suppose the initial three-mode quantum state is $\rho_{ACB}^0$, with covariance matrix $\gamma_{ACB}^{(0)}\ (=\gamma_{ACB})$, which can be written as
\begin{equation} 
    \gamma_{ACB}^{(0)}=\gamma_{ACB}=
    \begin{pmatrix} 
    V_{A}I_{2} & \phi_{AC}\,\sigma_{Z} & \kappa_{AB}^{(0)} \\ 
    \phi_{AC}\,\sigma_{Z} & V_CI_{2} & \psi _{CB}^{(0)}\\
    (\kappa_{AB}^{(0)})^T & (\psi _{CB}^{(0)})^T &\gamma_B^{(0)}
    \end{pmatrix},
\end{equation} 
where the upper-left $4\times 4$ sub-matrix $\gamma_{AC}$ is symmetric due to the intrinsic symmetry of the designed source, where 
\begin{equation}
    \begin{aligned}
    V_C=&(1-\eta_{BS})(V_A-1)+1, \\
    \phi _{AC}=&\sqrt{(1-\eta_{BS})\eta_A(V_A^2-1)},
    \end{aligned}
\end{equation}
and $I_2= (1\ 0\ ;\ 0\ 1)$, $\sigma_Z=(1\ 0\ ;\ 0\ -1)$. The remaining sub-matrices are estimated using the measurement data, resulting in an unknown matrix $\kappa_{AB}$, and two non-symmetric matrices $\psi_{CB}$ and $\gamma_{B}$,
\begin{equation}
    \begin{aligned}
    \kappa_{AB}^{(0)}&=(\kappa_{11}\ \kappa_{12}\ ;\ \kappa_{21}\ \kappa_{22}),\\
    \psi _{CB}^{(0)}&=(\phi_{11}\ \phi_{12}\ ;\ \phi_{21}\ \phi_{22}),\\
    \gamma_B^{(0)}&=(b_{11}\ b_{12}\ ;\ b_{12}\ b_{22}).
    \end{aligned}
\end{equation}
Then \(\gamma_{ACB}^{(0)}\) is symmetrized by the three steps as below:

\begin{figure}
    \centering
    \includegraphics[width= 7.5 cm]{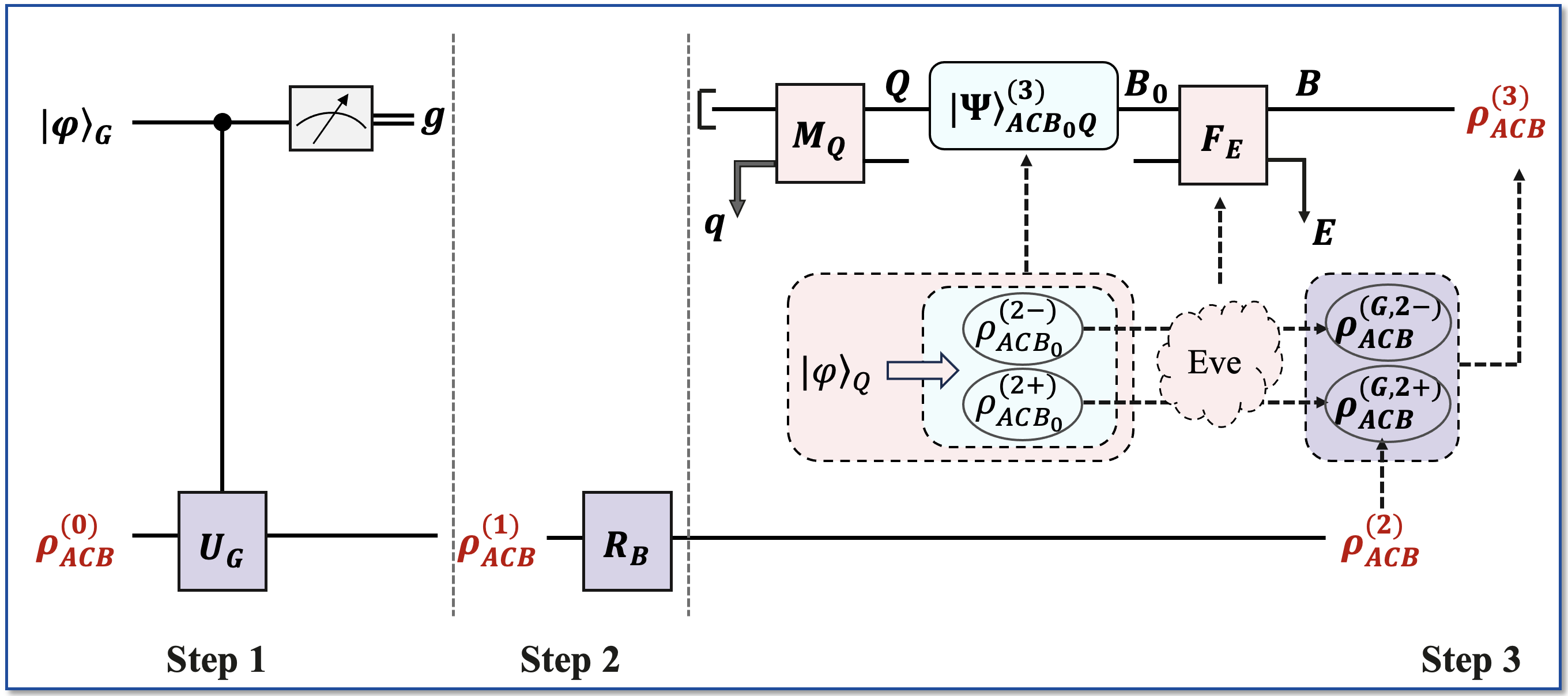}
    \caption{Covariance matrix diagonalization procedure. \textbf{Step 1:} Controlled rotation. Apply a controlled operation \(U_G\) to the state \(|\Psi\rangle_{ACB}^0\) conditioned on the control state \(|\varphi\rangle _G\) to obtain \(|\Psi\rangle_{ACB}^1\), and then \(|\varphi\rangle _G\) is measured to obtain the outcome \(g\). \textbf{Step 2:} Local rotation. Perform a local operation $R_B$ on system $B$. \textbf{Step 3:} State mixture. The two states are mixed according to $\big(\rho_{ACB}^{(G,2+)}+\rho_{ACB}^{(G,2-)}\big)\big/2$.
    }\label{app:Fig:SymStep}
\end{figure}

\textbf{Step 1:}\textit{ Controlled rotation.}
The first step symmetrizes the block $\gamma_B^{(0)}$. We introduce an auxiliary system $G$ as a control quantum state $|\varphi \rangle_G = \sqrt{1/2}(|0 \rangle_G+|1 \rangle_G)$, whose controlled operation $U_G=|0\rangle\langle0|_G\otimes I_{ABC} 
+|1\rangle\langle1|_G\otimes R_A\!\left(\tfrac{\pi}{2}\right)\otimes R_{BC}\!\left(-\tfrac{\pi}{2}\right)$ is applied to the target state $\rho_{ACB}^{(0)}$.
Then the global state becomes 
\begin{equation}
    \begin{aligned}
        \rho_{ACBG}^{(1)}=U_G\big(\rho_{ACB}^{(0)}\otimes|\varphi\rangle_G \langle \varphi|\big)U_G^{\dagger}.
    \end{aligned}
\end{equation} 
If \(|\varphi \rangle_G=|0\rangle_G\), the target state $\rho_{ACB}^{(0)}$ is left alone; if \(|\varphi \rangle_G=|1\rangle_G\), mode \(A\) is rotated by \(90^\circ\) while \(B\) and \(C\) are rotated by \(-90^\circ\), yielding the state $\rho_{ACB}^{\mathrm{(0R)}}$ with covariance matrix $\gamma_{ACB}^{\mathrm{(0R)}}$. After a quantum measurement on $G$, we can get the desired state $\rho_{ACB}^{(1)}$ and its covariance matrix
\begin{equation}
   \rho_{ACB}^{(1)}=\frac{\rho_{ACB}^{(0)}+\rho_{ACB}^{\mathrm{(0R)}}}{2},\ \gamma_{ACB}^{(1)}=\frac{\gamma_{ACB}^{(0)}+\gamma_{ACB}^{\mathrm{(0R)}}}{2},
\end{equation} 
where the sub-matrix $\gamma_{B}^{(1)}$ takes a symmetric form as 
\begin{equation}
   \gamma_{B}^{(1)}=V_BI_2=
   \begin{pmatrix}
    \frac{b_{11}+b_{22}}{2} & 0 \\ 
    0 & \frac{b_{11}+b_{22}}{2}
   \end{pmatrix}.
\end{equation} 
The remaining sub-matrices are
\begin{equation}
\kappa_{AB}^{(1)}=\begin{pmatrix}
\kappa^{(1)} & \Delta\kappa^{(1)} \\
\Delta\kappa^{(1)} & -\kappa^{(1)}
\end{pmatrix},
\ 
\psi _{CB}^{(1)}=\begin{pmatrix}
\phi^{(1)} & \Delta\phi^{(1)} \\
-\Delta\phi^{(1)} & \phi^{(1)}
\end{pmatrix},
\end{equation}
where $\kappa^{(1)}=(\kappa_{11}-\kappa_{22})/2$, $\Delta\kappa^{(1)}=(\kappa_{12}+\kappa_{21})/2$, $\phi^{(1)}=(\phi_{11}+\phi_{22})/2$, $\Delta\phi^{(1)}=(\phi_{12}-\phi_{21})/2$. It's obvious that $K(\rho_{ACB}^{(0)})=K(\rho_{ACB}^{\mathrm{(0R)}})$ because the rotation is a unitary operation. So based on the convexity of SKR, we have $K\big(\rho_{ACB}^{(1)}\big) \le K\big(\rho^{(0)}_{ACB}\big)$.

\textbf{Step 2:}\textit{ Local rotation.}
This step symmetrizes the block $\psi _{CB}^{(2)}$. A local unitary operation $R_B$ with a symplectic matrix $X_B=(\cos \theta\ -\sin \theta\ ;\ \sin \theta\ \cos \theta)$ is applied to mode $B$, yielding
\begin{equation}
    \begin{aligned}
          \rho_{ACB}^{(2)}=&(I_{AC} \otimes R_B)\rho_{ACB}^{(1)}(I_{AC} \otimes R_B)^{\dagger },\\ 
          \gamma_{ACB}^{(2)}=&(I_{4} \oplus X_B)\gamma_{ACB}^{(1)}(I_{4} \oplus X_B)^T.
    \end{aligned}
\end{equation}
We let $\tan \theta= -(\Delta\phi^{(1)})/(\phi^{(1)})$, then
\begin{equation}
    \begin{aligned}
        &\psi _{CB}^{(2)}=\psi _{CB}^{(1)}X_B^T=\phi_{CB}I_2 = \\
     &\begin{pmatrix}
    \sqrt{(\phi^{(1)})^2 + (\Delta\phi^{(1)})^2} & 0 \\ 
    0 & \sqrt{(\phi^{(1)})^2 + (\Delta\phi^{(1)})^2}
   \end{pmatrix}.
    \end{aligned}
\end{equation}
Similarly,
\begin{equation}
  \kappa _{AB}^{(2)}=\kappa _{AB}^{(1)}X_B^T=\begin{pmatrix}
\kappa^{(2)} & \Delta\kappa^{(2)} \\
\Delta\kappa^{(2)} & -\kappa^{(2)}
\end{pmatrix},  
\end{equation} 
where $\kappa^{(2)}=\kappa^{(1)} \cos \theta - \Delta\kappa^{(1)} \sin \theta$, and $\Delta\kappa^{(2)}=\kappa^{(1)} \sin \theta + \Delta\kappa^{(1)} \cos \theta$.
This unitary operation does not change SKR, therefore we have $K\big(\rho_{ACB}^{(2)}\big) = K\big(\rho_{ACB}^{(1)}\big)$.

\textbf{Step 3:}\textit{ State mixture.}
This step is to symmetrize the last block $\kappa _{AB}^{(2)}$. 
Let \(\rho_{ACB}^{(G,2+)}\) denote the Gaussian state sharing the same covariance matrix \(\gamma_{ACB}^{(2+)}\) (\(=\gamma_{ACB}^{(2)}\)) with \(\rho_{ACB}^{(2)}\). Define another Gaussian state \(\rho_{ACB}^{(G,2-)}\) associated with the covariance matrix \(\gamma_{ACB}^{(2-)}\), obtained from \(\gamma_{ACB}^{(2+)}\) by replacing \(\Delta\kappa^{(2)}\) with \(-\Delta\kappa^{(2)}\) while leaving all other entries unchanged. The corresponding symmetrized state and covariance matrix are
\begin{equation}
\rho_{ACB}^{\mathrm{sym}}=
\frac{\rho_{ACB}^{(G,2+)}+\rho_{ACB}^{(G,2-)}}{2},
\quad
\gamma_{ACB}^{\mathrm{sym}}=
\frac{\gamma_{ACB}^{(2+)}+\gamma_{ACB}^{(2-)}}{2},
\end{equation}
where the sub-matrix $\kappa _{AB}^{\mathrm{sym}}$ is symmetric and
\begin{equation}
    \kappa _{AB}^{\mathrm{sym}}=\kappa \sigma_Z= 
    \begin{pmatrix}
        \kappa^{(2)} & 0 \\
        0 & -\kappa^{(2)}
    \end{pmatrix}.
\end{equation}
Following the convexity of SKR, it can be proved that $K\big(\rho_{ACB}^{(3)}\big) \le K\big(\rho_{ACB}^{(2)}\big)$. 
Since \(\gamma_{ACB}^{(2+)}\) and \(\gamma_{ACB}^{(2-)}\) have the same symplectic spectra, the corresponding Gaussian states satisfy $K\left(\rho_{ACB}^{(G,2+)}\right)=K\left(\rho_{ACB}^{(G,2-)}\right)$.
Following the convexity of SKR and the Gaussian optimality theorem, $K\big(\rho_{ACB}^{\mathrm{sym}}\big) \le K\big(\rho_{ACB}^{(G,2+)}\big) \le K\big(\rho_{ACB}^{(2)}\big)$.

To obtain the mixed state, we first equally mix two entangled sources. Consider a physical source state \(|\Psi\rangle_{ACB_0}^{(2+)}\), which evolves into \(\rho_{ACB}^{(2)}\) after transmission through the quantum channel controlled by Eve. By Gaussian optimality, \(\rho_{ACB}^{(2)}\) can be replaced in the security analysis by the Gaussian state \(\rho_{ACB}^{(G,2+)}\) sharing the same covariance matrix \(\gamma_{ACB}^{(2+)}\). 
Since \(\gamma_{ACB}^{(2+)}\) and \(\gamma_{ACB}^{(2-)}\) have the same symplectic eigenvalues, Williamson's theorem \cite{app:GaussianQuantumInformation} guarantees that they are related by a symplectic transformation. Therefore, \(\rho_{ACB}^{(G,2-)}\) can be obtained from \(\rho_{ACB}^{(G,2+)}\) through the corresponding Gaussian unitary transformation and is also physical. Consequently, there exists a physical source state \(|\Psi\rangle_{ACB_0}^{(2-)}\), whose Gaussian counterpart after transmission is \(\rho_{ACB}^{(G,2-)}\).

Next, introduce another auxiliary system \(Q\), which can purify the entire system and yield 
$|\Psi\rangle_{ACB_0Q}^{(3)}=\frac{1}{\sqrt{2}}\bigl(|\Psi\rangle_{ACB_0}^{(2+)}|0\rangle_Q+|\Psi\rangle_{ACB_0}^{(2-)}|1\rangle_Q\bigr)$.
Then \(Q\) is measured and the result $q$ is published, with the result $q=|0\rangle_Q$ projecting the subsystem \(ACB_0\) onto \(|\Psi\rangle_{ACB_0}^{(2+)}\), and $q=|1\rangle_Q$ projecting it onto $|\Psi\rangle_{ACB_0}^{(2-)}$. After the measurement and Eve's attack, we can get the mixed state.

Ultimately, we obtain a scaling chain of SKR 
\begin{equation}
        K\big(\rho_{ACB}^{\mathrm{sym}}\big) \le K\big(\rho_{ACB}^{(2)}\big)
        =K\big(\rho_{ACB}^{(1)}\big) \le K\big(\rho^{(0)}_{ACB}\big),
\end{equation}
which demonstrates that we can safely calculate the SKR with a symmetric matrix, thus reducing the problem to solving for the single parameter $\kappa$. 

The above method is not limited to the case where the block $\gamma_{AC}$ is already in the symmetric form. More generally, an arbitrary two-mode Gaussian state can always be transformed into the standard form by local linear unitary Bogoliubov operations (LLUBOs) \cite{app:duan2000inseparability}. Since LLUBOs are local Gaussian unitary transformations, they leave the SKR invariant. Then the proposed three-step symmetrization procedure can be directly applied to general three-mode covariance matrices.

\subsection{Analytical solution for the range of $\kappa$}\label{app:kappa}
After we symmetrize the three-mode covariance matrix $\gamma_{ACB}$ and get $\gamma_{ACB}^{\mathrm{sym}}$, we reduce the problem to solving the single parameter $\kappa$. Applying the uncertainty-principle constraint $\gamma_{ACB}^{\mathrm{sym}}+i\Omega \ge 0$, the tightest feasible interval for $\kappa$ is obtained from the sixth-order principal minor, yielding $\kappa \in \mathcal{F}_{\kappa}=[\bar{\kappa}-R, \bar{\kappa}+R]$, where $\bar{\kappa} = \sqrt{T_{cb}\,\eta_{A}\,(1-\eta_{BS})\,\big(V_{A}^{2}-1\big)}$, and $R = \sqrt{T_{cb}\,\varepsilon_{cb}\,(1-\eta_{A})\,(V_{A}+1)}$.
The parameters \(T_{cb}\) and \(\varepsilon_{cb}\) are defined from the symmetrized two-mode covariance matrix \(\gamma_{CB}^{\mathrm{sym}}\) of modes \(B\) and \(C\) to provide an equivalent description of the channel transmittance and excess noise, where
\begin{equation}\label{app:eq:gamma_CB}
\gamma_{CB}^{\mathrm{sym}}=\begin{pmatrix} V_C I_2 & \phi_{CB} I_2 \\ \phi_{CB} I_2 & V_B I_2 \end{pmatrix}.
\end{equation}

For a matrix having the following form (assuming \(a>b\))
\begin{equation}\label{app:eq:gamma}
\gamma=\begin{pmatrix} aI_2 & cI_2 \\ cI_2 & bI_2 \end{pmatrix},
\end{equation}
the uncertainty principle requires $\det\bigl(\gamma+i\Omega\bigr)\ge 0$, which can be solved as
\begin{equation}\label{app:eq:c_upper}
c^2\le (a-1)(b-1)\le (a-1)^2.
\end{equation}
So we can define the channel transmittance and excess noise as
\begin{equation}\label{app:eq:T_def}
T_{\mathrm{ch}}=\frac{c^2}{(a-1)^2},\quad T\le 1, \quad \varepsilon_{\mathrm{ch}}=\frac{b-1}{T}-(a-1).
\end{equation}

Similarly, for the two-mode covariance matrix \(\gamma_{CB}^{\mathrm{sym}}\), we can define
\begin{equation}\label{app:eq:Tcb_epscb}
T_{cb}=\frac{\phi_{CB}^2}{(V_C-1)^2},\qquad
\varepsilon_{cb}=\frac{V_B-1}{T_{cb}}-(V_C-1).
\end{equation}
With these two parameters, we can obtain the analytical expression for the range of $\kappa$ as above. 

If we use the channel transmittance $T$ and excess noise $\varepsilon$ defined in Fig. \ref{app:EB}, as a linear channel model for performance simulation, then $\bar{\kappa}$ and $R$ can be re-expressed as $\bar{\kappa} = \sqrt{T\eta_{A}\eta_{BS}\,\big(V_{A}^{2}-1\big)}$, and $R = \sqrt{T\varepsilon(1-\eta_{A})(V_{A}+1)}$. And the correspondence between the parameters under these two definitions is
\begin{equation}
T_{cb}=\frac{\eta_{BS}}{1-\eta_{BS}}T,\qquad
\varepsilon_{cb}=\frac{1-\eta_{BS}}{\eta_{BS}}\varepsilon.
\end{equation}

As for finding an analytical solution for $\kappa^{\mathrm{worst}}$ that maximizes $S_{BE}^G$, the problem is quite complex. Here, we present an avenue for exploration. 
Treating \(\kappa\) as a perturbative parameter provides a possible route to characterize the local dependence of \(S_{BE}^G\) on \(\kappa\). The symplectic eigenvalues can be expanded perturbatively around a reference point and substituted into \(S_{BE}^G\) through the chain rule. The resulting expansion can then be used to analyze the stationary condition \(dS_{BE}^G/d\kappa=0\), yielding an approximate analytical equation for the stationary points of \(S_{BE}^G\). However, since the perturbative coefficients depend on the full symplectic spectrum of the covariance matrix, obtaining a general closed-form expression for the maximizing value of \(\kappa\) remains highly challenging. Developing such an analytical solution is therefore left for future investigation.
\begin{figure}[t]
    \includegraphics[width= 8 cm]{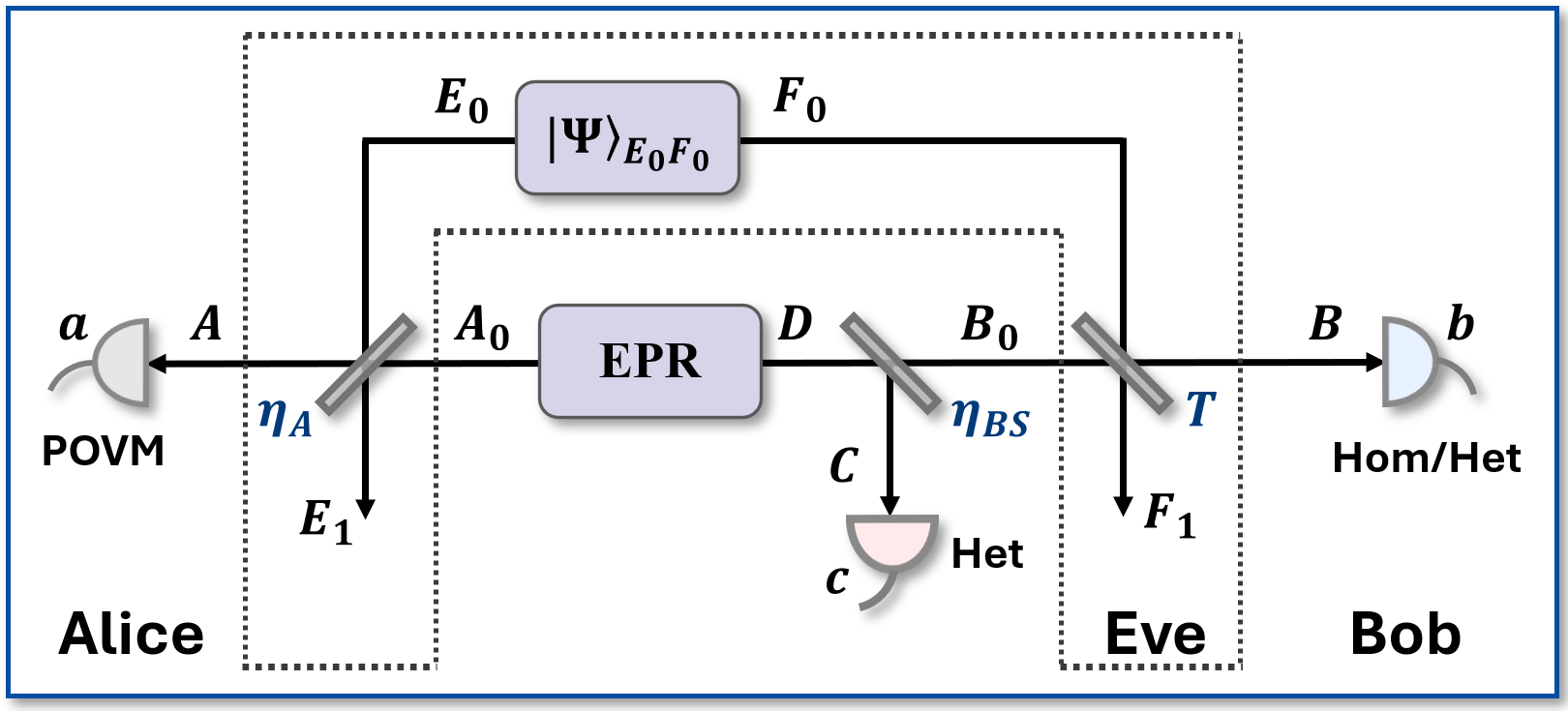}
    \caption{The attack model for the discrete-modulated protocol, in which the non-ideality of the discrete-modulated source is modeled as Eve's attack. 
    Alice prepares an Einstein-Podolski-Rosen (EPR) source and mode $D$ is divided by a beam splitter with transmittance $\eta_{BS}$ and get modes $C$ and $B_0$. Eve's attack is modeled as a two-mode state $|\Psi\rangle_{E_0F_0}$, where mode $E_0$ is coupled into the system via a beam splitter with transmittance $\eta_A$, mode $F_0$ is coupled into the system via a beam splitter with transmittance $T$. 
    }\label{app:Attackmodel}
\end{figure}

\subsection{Attack model}\label{app:model}

In this subsection, we provide a model to explain the unusual phenomenon that SKR exhibits nonlinear variations within the interval $\mathcal{F}_{\kappa}$. The model is presented in Fig. \ref{app:Attackmodel}, where non-ideality of the discrete-modulated source is modeled as an attack by Eve. Specifically, Alice prepares an Einstein-Podolski-Rosen (EPR) state, where mode $D$ is divided by a beam splitter with transmittance $\eta_{BS}$ and get modes $C$ and $B_0$. Eve's attack is modeled as a two-mode state $|\Psi\rangle_{E_0F_0}$, where mode $E_0$ is coupled into the system via a beam splitter with transmittance $\eta_A$, mode $F_0$ is coupled into the channel via a beam splitter with transmittance $T$. After Eve's attack, mode $A$ and mode $B$ are obtained. Then, Alice uses POVM for mode $A$ and heterodyne measurement for mode $C$. Mode $B$ is homodyne or heterodyne measured by Bob.


\begin{figure}[t]
    \centering
    \includegraphics[width=\linewidth]{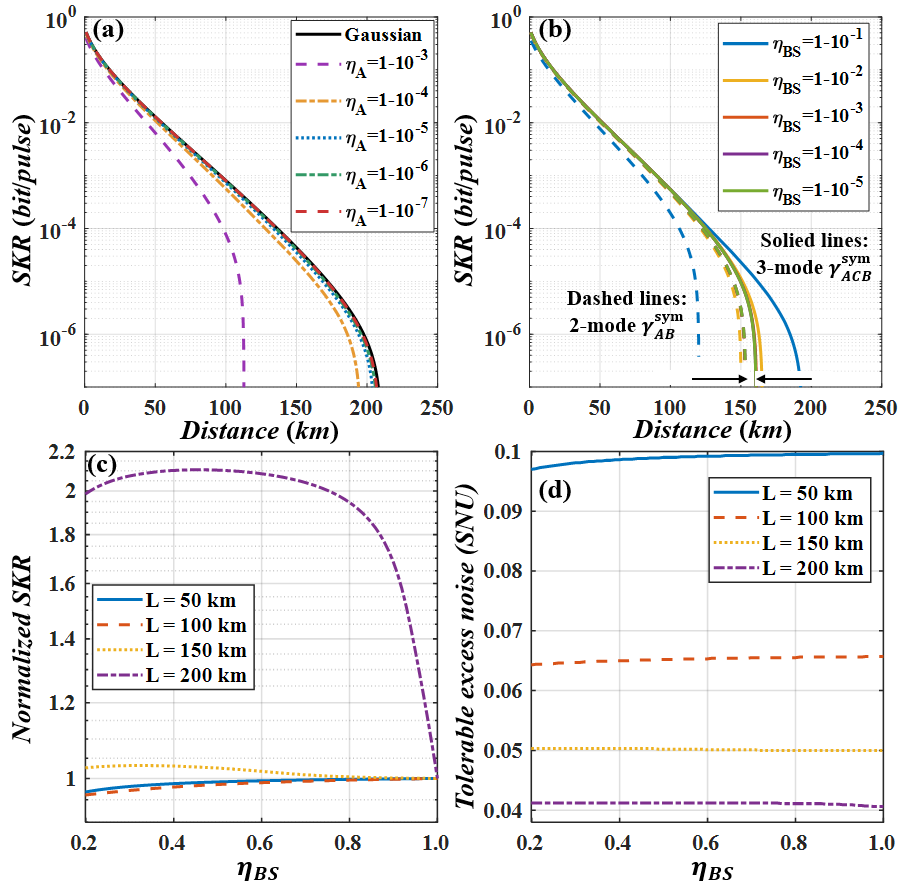}
    \caption{
        (a) SKR of the DM protocol for different $\eta_A$ compared with the ideal Gaussian protocol, where $V_A=5$ SNU and $\eta_{BS}=0.9$. (b) SKR based on 3-mode matrix $\gamma_{ACB}^{\mathrm{sym}}$ and 2-mode matrix $\gamma_{AB}^{\mathrm{sym}}$ with different $\eta_{BS}$. (c) Normalized SKR versus \(\eta_{BS}\) under different transmission distances, where the SKR is normalized to its value at \(\eta_{BS}=1-10^{-5}\). (d) Tolerable excess noise versus \(\eta_{BS}\) under different distances.
    Simulation parameters: reconciliation efficiency $\beta=0.95$, excess noise $\varepsilon=0.04$ SNU, detection efficiency $\eta=0.6$, electronic noise $\nu_{ele}=0.1$ SNU. $V_M=4$ and $\eta_A=1-10^{-4}$ for (b), $V_M=4$ and $\eta_A=1-10^{-6}$ for (c) and (d).
    }\label{app:Simulation1}
\end{figure}

According to this model, we can write the relationships between the different modes as follows:
\begin{equation}
    \begin{aligned}
        A =& \sqrt{\eta_A}A_0+\sqrt{1-\eta_A}E_0, \\
        B_0 =&\sqrt{\eta_{BS}}D_0+\sqrt{1-\eta_{BS}}N, \\
        B=& \sqrt{T}B_0+\sqrt{1-T}F_0,
    \end{aligned}
\end{equation}
where $N$ represents the vacuum state with variance $V_N=1$. The variance of mode $E_0$ is $V_{E_0}=V_{A}$. Mode $F_0$ represents the excess noise injected into the channel with variance $V_{F_0}=1+T\varepsilon /(1-T)$.

Thus, the covariance between mode $A$ and mode $B$ can be numerically calculated as 
\begin{equation}
    \begin{aligned}
        \phi_{AB} =& \sqrt{\eta_AT}\langle A_0B_0\rangle+\sqrt{1-\eta_A}\sqrt{1-T}\langle E_0F_0\rangle, \\
        =&\sqrt{\eta_AT\eta_{BS}}\sqrt{V_A^2-1}+\sqrt{1-\eta_A}\sqrt{1-T}\langle E_0F_0\rangle.
    \end{aligned}
\end{equation}
According to the uncertainty principle, we have
\begin{equation}
    \begin{aligned}
        |\langle E_0F_0\rangle| \le & \sqrt{(V_A+1)(V_{F_0}-1)} \\
        = & \sqrt{(V_A+1)\frac{T\varepsilon}{1-T}},
    \end{aligned}
\end{equation}
When the covariance $\langle E_0F_0\rangle $ takes its minimum negative value, the two-mode state $|\Psi\rangle_{E_0F_0}$ is an ideal entangled state. At this point, the covariance between modes $A$ and $B$ is
\begin{equation}
        \phi_{AB}^{\mathrm{min}} =\sqrt{\eta_AT\eta_{BS}}\sqrt{V_A^2-1}-\sqrt{(1-\eta_A)T\varepsilon (V_A+1)}.
\end{equation}

Interestingly, the value of $\phi_{AB}^{\mathrm{min}}$ is exactly equal to the left boundary of the interval, i.e., $\phi_{AB}^{\mathrm{min}}=\bar{\kappa}-R$. 
This provides some insight on the non-monotonic behavior of the secret key rate, which is not necessarily minimized at the left boundary under different conditions.
As $\phi_{AB}$ changes from the mean value to the left boundary, Eve's two mode state $\rho_{E_0F_0}$ evolves from a mixed tensor product state to an entangled state, with its degree of entanglement reaching a maximum at the left boundary. While not rigorously proven, numerical observations suggest that, when SKR exhibits a non-monotonic dependence on $\kappa$, the maximum value of $S_{BE}^G$ happens near the point where $\rho_{E_0F_0}$ becomes entangled.
Since multipartite entanglement exhibits complex and potentially non-monotonic behavior, Eve's accessible information may not be a monotonic function of $\kappa$. The underlying mechanism remains unclear and deserves further investigation.

\subsection{Simulations}\label{app:Simulation}


We present some additional simulated asymptotic key rates in this subsection. 
Firstly, a key factor affecting SKR is the design of entangled source. As shown in Fig. \ref{app:Simulation1} (a), for a fixed mode variance $V_A=5$, larger $\eta_A$ yields better performance. The results further indicate that $\eta_A$ should be at least on the order of $1-10^{-4}$ to avoid significant performance degradation. 
Moreover, when $\eta_A$ is on the order of $1-10^{-7}$, the SKR becomes nearly indistinguishable from that of the ideal Gaussian protocol. Such a source can be realized using no more than 256 constellation points.
Fig.~\ref{app:Simulation1} (b) shows the SKR under different \(\eta_{BS}\) calculated from the 3-mode matrix $\gamma_{ACB}^{\mathrm{sym}}$ and 2-mode matrix $\gamma_{AB}^{\mathrm{sym}}$, where the same transmitted states are considered with a 64-QAM constellation and modulation variance \(V_M=4\). Overall, the SKR from $\gamma_{AB}^{\mathrm{sym}}$ is slightly lower than that from $\gamma_{ACB}^{\mathrm{sym}}$. As \(\eta_{BS}\) increases, the SKR of the two-mode scheme gradually increases, while for the three-mode scheme, the SKR decreases at long distances but increases at short distances. Consequently, the discrepancy between the SKRs gradually diminishes, and the corresponding curves eventually converge to a same one.
Fig.~\ref{app:Simulation1} (c) and (d) show the normalized SKR and the maximum tolerable excess noise as functions of \(\eta_{BS}\) under different transmission distances. Both quantities exhibit similar behavior: increasing \(\eta_{BS}\) generally improves the performance, particularly in the short-distance regime, while the improvement gradually saturates as \(\eta_{BS}\rightarrow 1\). At longer distances, the enhancement becomes weaker and may even slightly decrease in the high-\(\eta_{BS}\) regime. This reflects the trade-off introduced by the beam-splitter transmittance.

\balance


\end{document}